\def\revtex{1}

\ifx\revtex\undefined

\documentclass[review]{elsarticle}

\usepackage{lineno,hyperref}
\modulolinenumbers[5]

\journal{Theoretical Computer Science}

\usepackage[dvipsnames]{xcolor}
\usepackage{booktabs}
\usepackage{amssymb}
\usepackage{amsmath}
\usepackage{mathbbol} %%%% for \mathds{1}
\usepackage{tikz}
\usetikzlibrary{calc,decorations.pathreplacing,decorations.markings,positioning,shapes}

\begin{document}

\begin{frontmatter}

\title{Varieties of contextuality based on probability and structural nonembeddability}

\author{Karl Svozil}
\address{Institute for Theoretical Physics, TU Wien, Wiedner Hauptstrasse 8-10/136, 1040 Vienna,  Austria}

\ead[url]{tph.tuwien.ac.at/~svozil}

\cortext[mycorrespondingauthor]{Karl Svozil}
\ead{svozil@tuwien.ac.at}


\begin{abstract}
Different analytic notions of contextuality fall into two major groups: probabilistic and strong notions of contextuality. Kochen and Specker's Theorem~0~\cite{kochen1} is a demarcation criterion for differentiating between those groups. Whereas probabilistic contextuality still allows classical models, albeit with nonclassical probabilities, the logico-algebraic ``strong'' form of contextuality characterizes collections of quantum observables that have no faithfully embedding into (extended) Boolean algebras. Both forms indicate a classical in- or under-determination that can be termed ``value indefinite'' and formalized by partial functions of theoretical computer sciences.
\end{abstract}


% Author Orchid ID: enter ID or remove command
%\newcommand{\orcidauthorA}{0000-0001-6554-2802} % Add \orcidA{} behind the author's name
%\newcommand{\orcidauthorB}{0000-0000-0000-000X} % Add \orcidB{} behind the author's name

%\PACS{03.65.Ca, 02.50.-r, 02.10.-v, 03.65.Aa, 03.67.Ac, 03.65.Ud}

\end{frontmatter}

\linenumbers

\else
\documentclass[%
  reprint,
  twocolumn,
 %superscriptaddress,
 %groupedaddress,
 %unsortedaddress,
 %runinaddress,
 %frontmatterverbose,
 % preprint,
 showpacs,
 showkeys,
 preprintnumbers,
  nofootinbib,
 %nobibnotes,
 %bibnotes,
 amsmath,amssymb,
 aps,
 % prl,
 pra,
 % prb,
 % rmp,
 %prstab,
 %prstper,
  longbibliography,
 %floatfix,
 %lengthcheck,%
 ]{revtex4-2}

\usepackage{booktabs}

\usepackage[dvipsnames]{xcolor}

\usepackage{mathptmx}% http://ctan.org/pkg/mathptmx

\usepackage{amssymb,amsthm,amsmath,bm}

\usepackage{tikz}
\usetikzlibrary{calc,decorations.pathreplacing,decorations.markings,positioning,shapes,snakes}
%\usetikzlibrary{calc,decorations.pathreplacing,decorations.markings,positioning,shapes,snakes,external}
%\tikzexternalize

\usepackage[breaklinks=true,colorlinks=true,anchorcolor=blue,citecolor=blue,filecolor=blue,menucolor=blue,pagecolor=blue,urlcolor=blue,linkcolor=blue]{hyperref}
\usepackage{graphicx}% Include figure files
\usepackage{url}

%%%%%%%%%%%%%%%%%%%%%%%%%%%%%
\usepackage{iftex}
\ifxetex
%
% XeLaTeX
%
\usepackage{fontspec}
\usepackage{fontspec}
\setmainfont{Garamond}
\setsansfont{Garamond}
\fi
%%%%%%%%%%%%%%%%%%%%%%%%%%%%%

\usepackage{mathbbol} %%%% for \mathds{1}

\begin{document}

\title{Varieties of contextuality based on probability and structural nonembeddability}

\author{Karl Svozil}
\email{svozil@tuwien.ac.at}
\homepage{http://tph.tuwien.ac.at/~svozil}

\affiliation{Institute for Theoretical Physics,
TU Wien,
Wiedner Hauptstrasse 8-10/136,
1040 Vienna,  Austria}

\date{\today}

\begin{abstract}
Different analytic notions of contextuality fall into two major groups: probabilistic and strong notions of contextuality. Kochen and Specker's demarcation Theorem~0~\cite{kochen1} is a criterion for differentiating between those groups. Whereas probabilistic contextuality still allows classical models, albeit with nonclassical probabilities, the logico-algebraic ``strong'' form of contextuality characterizes collections of quantum observables that have no faithfully embedding into (extended) Boolean algebras. Both forms indicate a classical in- or under-determination that can be termed ``value indefinite'' and formalized by partial functions of theoretical computer sciences.
\end{abstract}


\maketitle

\fi

\section{Types of quantum contextuality}

The main point of this paper is that there are at least two main types of contextuality: the first notion is based upon nonclassical phenomenology,
and in particular, on nonclassical probabilities contradicting Boole's conditions of physical experience~\cite{Boole-62}.
The second type goes beyond this,
and is based upon the absence of a classical logico-algebraic structure in terms of which the respective observables could be re-interpreted.
Formally this amounts to nonembaddability into Boolean algebras by classical means; that is, in formal terms,
by two-valued measures interpretable as classical truth assignments. The strongest form of this latter logico-algebraic contextuality
is the total absence of any such classical truth assignment~\cite{kochen1}.
Thereby, it is important to keep in mind that the mere existence of any such classical truth assignment
is necessary but not sufficient to ensure classical representability or embeddability:
indeed, even if there is an apparent ``abundance'' of classical truth assignments, these might
not provide sufficient means to classically embed a collection of quantum observables.

Early synthetic conceptions of contextuality emerged from
insights into the entangled complexion~\cite{schrodinger-gwsidqm2} of physical properties
retrieved from quantum measurements. As expressed by Bohr~\cite{bohr-1949}:
{``the impossibility of any sharp separation between the behavior of atomic
objects and the interaction with the measuring instruments which serve to define the conditions
under which the phenomena appear.''}
This yields a {``conditionality of phenomena''}~\cite{Khrennikov2017,Jaeger2019} relative to a
{``a complex of conditions under which the measurement is performed''}~\cite{Khrennikov2009a,Khrennikov2009b}.

In this line of thought, observable phenomena appear not as isolated properties of the object,
but as signals from the object-measurement apparatus composite.
(Entanglement may even extend to the observer~\cite{london-Bauer-1939,london-Bauer-1983}.)
Indeed, if entanglement is involved, these signals are about the relational properties of the combined entangled system.
It makes no sense to refer to a well-defined property of the individual object alone~\cite{schrodinger-gwsidqm2,zeil-99}.
Therefore, one should be careful interpreting a statement such as Bell's observation~\cite{bell-66} that
{``the result of an observation may reasonably depend $\ldots$ on the complete disposition of the apparatus''.}
In general there is no deterministic, one-to-one correspondence, association, or translation between relevant
(counterfactual) well-defined individual properties of the constituents (one imagined~\cite{hertz-94} as ``object'')
of an entangled quantized system on the one hand,
and the signal resulting from observation of this entangled state on the other hand. Entanglement evades such an association
because the constituents of an entangled quantum state have no well-defined individuality.

Subsequent attempts to specify and quantify contextuality
have presumed that individual objective properties nevertheless exist even for quantized systems,
and that these properties do not depend on any kind of disposition of the measurement apparatus;
in particular not on some compatible observables that are measured simultaneously.
The latter assumption is usually referred to as ``noncontextual''.
The former assumption of the general existence of counterfactual properties or observables can be called omni-existence.
Omni-existence is the ``totality'' assertion that, although due to complementarity not all observables can be measured simultaneously, they are nevertheless value definite;
that is, some of them have a definite counterfactual value that can by no physical means be measured~\cite{specker-60}.

Historical attempts to prove contextuality have assumed both omni-existence and noncontextuality
(thereby disregarding earlier synthetic concepts of contextuality by entanglement mentioned earlier), and
have concentrated on the differences between classical and quantum predictions.
Thus, given  omni-existence, any empirically (falsifiable) discrepancies between classical and quantum predictions are interpreted to signify contextuality.
However, one has to be careful and keep in mind that, just because omni-existence is often but not always~\cite{2015-AnalyticKS} assumed for the sake of contradiction,
a violation of noncontextuality does not imply or suggest the existence of any contextual hidden variable model.
Contemporary interpretations of contextuality indicate the non-existence of a noncontextual model~\cite{cabello2021contextuality}.

Most commonly, experimental violations of Boole-Bell-type inequalities are identified with quantum contextuality~\cite{cabello:210401,cabello2021contextuality}.
Other empirical signatures of quantum contextuality are the experimental violations of ad hoc configurations
whose classical interpretation
(i)
either merely assume the omni-existence of unrestricted classical noncontextual value assignments that do not depend on the
complete disposition of the apparatus~\cite{cabello:210401};
(ii)
or, on preselected input, predict classical functional output that is violated in quantized systems~\cite{2018-minimalYIYS}.

Theoretical arguments against omni-existent noncontextual value assignments consider finite configurations of
observables forming intertwining contexts that have no consistent classical value assignments~\cite{specker-60}.
Here the term ``intertwined'' is understood as introduced by Gleason~\cite{Gleason}:
in higher than two dimensions the orthonormal bases identified with contexts can ``share'' common elements---they need not be ``isolated'', that is, disjoint.
This is not the place for a historic review but the literature indicates that
what is now known as the Kochen-Specker theorem~\cite[Theorem~1]{kochen1} has been
discussed~\cite{ZirlSchl-65,kamber65} as a direct consequence of Gleason's theorem~\cite{Gleason}.

The variety of contextual signifiers has resulted in a great semantic spread of notions of contextuality that threatens to obscure subtle differences
concerning the quality of anomaly.
Because the same collection of observables, taken from quasi-classical or quantum experimental configurations alike,
may still allow very different types of probability distributions.
The resulting differences in the prediction may be taken as signatures for contextuality.
But this is incomparable to configurations of observables that do not, by any classical means, support such probability distributions,
either because the existing classical value assignments that do not depend on the complete disposition of the apparatus
are too scarce to resolve the logico-algebraic structure of observables at hand,
or because this structure does not allow any such classical value assignment at all.

In what follows I, therefore, suggest refining the notion of contextuality by differentiating between two cases,
depending on whether the collection of observables
\begin{itemize}
\item[(i)] violates some constraints on classical probabilities but still allows a faithful embedding into an extended Boolean subalgebra, or
\item[(ii)] does not, by any classical means, allow any faithful embedding into some extended Boolean subalgebra.
\end{itemize}

For the sake of this analysis note that maximal collections of (finitely many) mutually co-measurable observables can be ``wrapped up'' into
contexts (or blocks or maximal observables) which, from a probabilistic and structural point of view, intrinsically behave classically.
The respective probability distributions are in accord with Kolmogorov's axioms. In particular,
the probabilities of mutually exclusive events add up~\cite{Gleason}.

\section{Nomenclature}

For quantum mechanics we fix a positive integer $n\ge 2$.
Let $O$ be a nonempty set of one-dimensional projection observables on the Hilbert space $\mathbb{C}^n$~\cite{halmos-vs}.
A set $C\subset O$ is a {\em context} of $O$ if $C$ is an orthonormal basis of $\mathbb{C}^n$.
An equivalent definition is in terms of the spectrum of a maximal operator containing one-dimensional projection operators
onto the subspaces spanned by the vectors in $C$.
In more general empirical logic terms, a context can be conceptualized by a set of mutually exclusive observables whose disjunction is a tautology.

Quantum mechanics, as well as partition logics~\cite{svozil-2001-eua} from a generalized urn~\cite{wright} or finite automata models~\cite{e-f-moore,schaller-96} featuring complementarity, allow
two or more distinct contexts
which, for more than two mutually exclusive outcomes per context, may intertwine in some observable(s)~\cite{svozil-2018-b}.
The remaining nonintertwining observables,
taken from different contexts, exhibit complementarity.
(One convenient and compact graphical representation depicts contexts as smooth lines,
and mutually exclusive elementary observables by points on these lines.)

As has been mentioned earlier, despite exhibiting complementarity,
a collection of contexts may still allow some quasi-classical interpretation in terms of
extreme cases.
Such value assignments can be formalized by dispersionless {\em two-valued states} $v(x) \in \{0,1\}$ (or, logically interpreted, ``false'' and ``true'')
that are binary functions of the respective observables $x \in X$ forming the contexts
which are additive $v(x \vee y)= v(x) + v(y)$ for mutually exclusive observables $x\wedge y=\emptyset$, and add up to 1
for all mutually exclusive observables within any context.
A weak and general formalization for quantized systems that allows partial functions and thus value indefiniteness
is in terms of admissibility~\cite{2015-AnalyticKS}:
Let $O$ be a set of one-dimensional projection observables and let $v: O \to \{0,1\}$ be a value assignment function.
Then $v$ is {\em admissible} if the following two conditions hold for every context $C$ formed by $O$:
\begin{itemize}
\item[(i)] exclusivity: if there exists an $x\in C$ with $v(x)=1$, then $v(x')=0$ for all $x'\in C \setminus \{x\}$;
\item[(ii)] completeness: if there exists an $x\in C$ with $v(x')=0$ for all $x'\in C\setminus \{x\}$, then $v(x)=1$.
\end{itemize}

If the observables on which the aforementioned collections of contexts are based are quantum,
then there is no guarantee that ``sufficiently many'' classical value assignments corresponding to dispersionless two-valued $\{0,1\}$-states exist.
Indeed, there are finite configurations of observables with
no such classical value assignment, or ones that cannot support ``sufficiently many'' classical value assignments to allow
embeddings preserving the respective logico-algebraic structure.

Already Kochen and Specker discussed these issues and presented a demarcation criterion~\cite[Theorem~0]{kochen1} that
utilizes the (in)separability of the underlying binary elementary propositions by classical value assignments:
A set of observables  $X$ forming a collection of contexts
is faithfully embeddable into an extended Boolean algebra if and only if these observables in $X$ contained in the respective contexts
support or allow a separating set of two-valued states $V = \{v_1, \ldots ,v_n\}$ such that,
for any two observables $x , y \in X$,
there exists some $v_i \in V$ for which $v_i (x) \neq v_i (y)$.
(In its extreme form the collection of observables $X$ support no two-valued measure, and $V=\emptyset$.)

If the observables in $X$ support a separating set of two-valued states $V$ then $V$ facilitates three constructions:
\begin{itemize}
\item[(i)]
It yields all classical probability distributions
in the form of a convex combination
$
P(x) = \sum_{i=1}^n\lambda_i v_i(x)$,
such that
$
\sum_{i=1}^n\lambda_i=1
$
and
$
\lambda_j > 0
$
for all
$
j\in \left\{1,\ldots,n\right\}
$.

\item[(ii)]
The values of the value assignments formalized by dispersionless two-valued states on, say, $k$ observables can be arranged in $k$-tuples.
These tuples can
be interpreted as extreme points or vertices of a compact convex subset of $\mathbb{R}^k$, a convex polytope,
that has an equivalent representation in terms of its hull; that is, as a set of half-spaces which are (in)equalities.
In the quantum physical realm, these inequalities are often referred to as Boole-Bell-type inequalities~\cite{froissart-81,pitowsky-86}.
We shall later encounter such a hull computation in deriving the Suppes-Zanotti inequalities~(\ref{2020-ex-SZI}).

\item[(iii)]
A complete set of $n$ two-valued states allows a representation as a partition logic
that explicitly represents a classical embedding into an extended Boolean algebra $2^n$~\cite{svozil-2001-eua}.
\end{itemize}

Therefore, as disclosed earlier, it is suggested to adopt Kochen and Specker's demarcation criterion of embeddability for a refined definition of contextuality:
One could speak of {\em strong contextuality} if no classical representation of the respective observables, and also no classical probability distribution exists.
Strong contextuality always indicates
some ``essential'' scarcity of two-valued states associated with classical truth assignments;
a deficiency to supply sufficient ``structural information'' for a classical embedding of the respective quantum observables.
As will be discussed later such essential scarcity can be categorized and quantified by escalating levels of ``rarity'', a ``shortage''
of elements of the set of two-valued states that may become nonseparating, nonunital, or in its strongest form empty.

The weaker
{\em probabilistic contextuality}, while featuring complementarity (because more than one context is involved),
allows all kinds of classically embeddable collections of observables, as well as classical probability distributions---if
only the
probability distribution in some way differs from global classical Kolmogorovian probabilities that are not restricted to local contexts.
That is, the observables still allow classical probability
distributions---as well as ``hidden variables'' in terms of the ``larger'' extended Boolean algebra in which the observables can be homomorphically embedded---but
those probabilities are not realized by the  probabilistic contextual  systems at hand.

Why should one make such a distinction from a physical point of view?
Because if there do not exist separating sets of two-valued states and thus no faithful classical embeddability
then the respective structures do not any longer support classical probability distributions;
and also no ``hidden variables'' in terms of the ``larger'' extended Boolean algebra in which the observables can be homomorphically embedded.
This is quite different from any nonclassical probability distribution on an otherwise ``quasi-classical''  empirical logics~\cite{Foulis1976}
containing complementary observables that can still be imbedded into ``larger'' Boolean algebras.

For physical realizations we refer to Wright's generalized urn models~\cite{wright} or the initial state identification problem
for finite automata~\cite{e-f-moore,schaller-96}, amounting to partition logics~\cite{svozil-2001-eua}.
The resulting empirical logics~\cite{Foulis1976} are identical to a variety of quantum propositional structures~\cite{birkhoff-36}
These logics support both classical as well as quantum probabilities.
Which probability distribution needs to be chosen depends solely on the respective physical realization~\cite{svozil-2018-b}.
In particular, there is no structural logical distinction; therefore, classical ``hidden variables'' cannot be ruled out,
the difference is in the classical-versus-quantum performance: the respective quantum contextuality is probabilistic.

In what follows several quantum mechanical examples of probabilistic contextuality in $n$--dimensional Hilbert space will be enumerated.
To avoid unnecessary redundancies they are mentioned with a reference to the concrete computation.
Often two-valued states will be rewritten in terms of the
expectation values of dichotomic outcomes $E \in \{-1,1\}$
by affine transformations---a multiplication followed by a subtraction---from classical value assignments encoded by two-valued states
$v \in \{0,1\}$:
$
E = 2 v - 1
$, or conversely,
$
v =  \frac{1}{2}\left( E + 1 \right)
$.
In quantum mechanics and Hilbert spaces of dimension greater than one, $E$ generalizes to a unitary Householder transformation
$
\textsf{\textbf{E}}_{\bf x}
=
\mathbb{1}- 2   {\bf x}^\dagger {\bf x}
$,
where ${\bf x}$ is a unit vector and $\dagger$ represents the Hermitian adjoint (aka conjugate).
The resulting eigensystem of $\textsf{\textbf{E}}_{\bf x}$ has eigenvalues $\pm 1$:
\begin{itemize}
\item[$-1$:]
${\bf x}$ is an eigenvector of    $\textsf{\textbf{E}}_{\bf x}$ with eigenvalue $-1$.
\item[$+1$:]
The remaining $n-1$ mutually orthogonal eigenvectors span the $n-1$ dimensional subspace orthogonal to ${\bf x}$.
Every vector in that subspace has eigenvalue $+1$. (For $n>2$ the spectrum is degenerate.)
\end{itemize}
For any context represented by some orthonormal basis
$
%{\frak B}=
\{  {\bf e}_1  ,
  {\bf e}_2  , \ldots ,  {\bf e}_n  \}$,
the product of the respective unitary Householder transformations is minus the identity; that is,
$
\textsf{\textbf{E}}_{{\bf e}_1} \textsf{\textbf{E}}_{{\bf e}_2} \cdots \textsf{\textbf{E}}_{{\bf e}_n}
=
-\mathbb{1}
$.

All such approaches take ``a bag of'' observables from quantum mechanics---some of them complementary---and force a classical interpretation upon them.
This is done in terms of classical value assignments formalized by two-valued states or expectations of binary observables.
Note that there exist classical empirical models featuring complementarity,
such as the ones meantioned earlier: Moore automata~\cite{e-f-moore,schaller-96}, or generalized urn models~\cite{wright},
both yielding partition logics~\cite{svozil-2001-eua,svozil-2018-b}.

\section{Examples of probabilistic contextuality}

In what follows we shall concentrate on two subtypes of  probabilistic contextuality;
one based on Boole's  ``conditions of physical experience''~\cite{Boole-62},
and another one on the functional behaviour derived from terminal vertices of gadget graphs.

\subsection{Boole-Bell type signatures of contextuality by hull computations}

As mentioned earlier the hull computation of the convex polytope
formed by vertices that represent the encoded classical value assignments of a given selection of observables
yields inequalities that are identified
with optimal Boole-Bell type inequalities~\cite{froissart-81,pitowsky-86}.
Violations of these inequalities
by quantum probabilities are interpreted as signifying contextuality relative to the assumptions discussed earlier,
in particular, omni-existence and noncontextuality.

For the sake of concrete examples of historic configurations used for Boole-Bell type inequalities
consider configurations with
\begin{itemize}
\item[(i)] isolated contexts with no common observable, such as
\begin{itemize}
\item[(i.SZ)]
Suppes-Zanotti inequalities from three observables~\cite{Suppes-81}, as discussed later;
\item[(i.CHSH)]
Clauser-Horne-Shimony-Holt (CHSH) inequalities from four observables in two groups, and their respective tensor products forming four isolated contexts~\cite{Suppes-81};
\item[(i.TPB)]
two-party Bell inequalities from finitely many observables, and their respective tensor products forming isolated contexts~\cite{Avis2005};
\end{itemize}
\item[(ii)]
intertwining contexts from three-or higher dimensional Hilbert spaces with common observables, such as
\begin{itemize}
\item[(ii.SB)]
for the Specker bug configuration~\cite{kochen2} that serves as graph theoretic true-implies-false gadget~\cite{svozil-2001-cesena};
\item[(ii.KCBS)]
inequalities from five cyclically connected contexts~\cite{Klyachko-2008}.
(The Bub-Stairs inequality~\cite{Bub-2009} on the same configuration is ad hoc and does not follow from a hull
computation but from a classical probability assessment.)
\end{itemize}
\end{itemize}

Suppes and Zanotti's result~\cite{Suppes-81,svozil-2020-ex} as well as other bounds on classical probabilities for different configurations of observables
still allow {\em single} instances of $\{-1,1\}$-value assignments, as the respective set of two-valued states is not empty (as for Kochen-Specker configurations).
But the classical probabilistic analysis of such configurations reveals that the bounds on classical probabilities
are violated by the quantum probabilities of analogous quantized systems.
This has been, for instance, pointed out by tabulations of classical  $\{-1,1\}$-value assignments
for the CHSH configuration~\cite{chsh} by Asher Peres~\cite{peres222}.
Further quantitative investigations into the ``amount'', or a measure, of probabilistic contextuality
in terms of enumerations and tabellations of classical value assignments have, for instance,
been studied in References~\cite{svozil_2010-pc09,svozil-2011-enough,Dzhafarov-2015,Dzhafarov-2017,KujalaDzhafarov-2019,Dzhafarov-19PhysRevA}.

Let us, for the sake of an explicit example, enumerate the classical value assignments in the Suppes-Zanotti configuration
which consists of measurements of an Einstein-Podolski-Rosen~\cite{epr} type configuration of two observables on one ``side'',
and one observable on the other ``side'' (CHSH uses a symmetric 2-2 configuration).
It therefore involves $2+1=3$ binary observables $X,Y,Z\in \{-1,+1\}$ associated with $\{-1,1\}$-value assignments
that can be used to form the second-order distributions
from the three second-order expectations
$\textsf{\textbf{E}}(X,Y) = XY$,
$\textsf{\textbf{E}}(X,Z) = XZ$,
$\textsf{\textbf{E}}(Y,Z) = YZ$,
obtained by just multiplying the binary observables from their three possible pairs, respectively.
Classically this amounts to $2^3=8$ value assignments tabulated in Table~\ref{2021-context-table-sz}.
\begin{table}%[H] %add [H] placement to break table across pages
\centering
%\begin{ruledtabular}
\begin{tabular}{c|ccc|ccccccccccccccccccc}
\toprule
\textbf{\#} & $X$ & $Y$ & $Z$ & $XY$ & $XZ$ & $YZ$ \\
\midrule
$v_1$ & $+1$ & $+1$ & $+1$ & $+1$ & $+1$ & $+1$ \\
$v_2$ & $+1$ & $+1$ & $-1$ & $+1$ & $-1$ & $-1$ \\
$v_3$ & $+1$ & $-1$ & $+1$ & $-1$ & $+1$ & $-1$ \\
$v_4$ & $-1$ & $+1$ & $+1$ & $-1$ & $-1$ & $+1$ \\
$v_5$ & $+1$ & $-1$ & $-1$ & $-1$ & $-1$ & $+1$ \\
$v_6$ & $-1$ & $+1$ & $-1$ & $-1$ & $+1$ & $-1$ \\
$v_7$ & $-1$ & $-1$ & $+1$ & $+1$ & $-1$ & $-1$ \\
$v_8$ & $-1$ & $-1$ & $-1$ & $+1$ & $+1$ & $+1$ \\
\bottomrule
\end{tabular}
%\end{ruledtabular}
 \caption{\label{2021-context-table-sz}  The eight  $\{-1,1\}$-value assignments of the Suppes-Zanotti configuration.}
 \end{table}
These eight classical value assignments can be used to generate all classical higher-order distributions~\cite{Garg1984}
by convex summation of these value assignments; in particular, the probabilities
\begin{equation}
\begin{split}
p(x) = \lambda_1 v_1(x) +\cdots + \lambda_n v_n(x),
\\
{\rm with} \;
\lambda_1 +\cdots + \lambda_n=1
\text{ and }
 \lambda_j  \ge  0
, \;
j\in \left\{1,\ldots,n\right\}
.
\end{split}
\end{equation}
The second-order distributions can be geometrically characterized by a convex polytope~\cite{froissart-81,pitowsky-86}
formed by the ``classical vertices'' whose coordinates are arranged in three-tuples
$\begin{pmatrix}XY,XZ,YZ\end{pmatrix}^\intercal$ ($\intercal$ indicates transposition)
with respect to the Cartesian standard basis of $\mathbb{R}^3$  are identified
with  the respective last three row entries of Table~\ref{2021-context-table-sz}:
in this case the four vertex vectors (the other four vectors are duplicates) are
$\begin{pmatrix} 1,1,1\end{pmatrix}^\intercal$,
$\begin{pmatrix} 1,-1,-1\end{pmatrix}^\intercal$,
$\begin{pmatrix} -1,1,-1\end{pmatrix}^\intercal$,
and
$\begin{pmatrix} -1,-1,1\end{pmatrix}^\intercal$.
The resulting convex polytope has an equivalent representation in terms of its hull, formed by its half-spaces~\cite{ziegler,Schrijver,Fukuda-techrep}.
In the case of the Suppes-Zanotti configuration~\cite{Suppes-81,Brody-1989,Khrennikov2020} the hull equations are
the Suppes-Zanotti inequalities; in particular, the four half-spaces described by the inequalities~\cite{svozil-2020-ex}
\begin{equation}
- 1  \le  \pm \textsf{\textbf{E}}(X,Y) \pm \textsf{\textbf{E}}(X,Z) \pm \textsf{\textbf{E}}(Y,Z) \le 1
.
\label{2020-ex-SZI}
\end{equation}

Quantization of the Suppes-Zanotti configuration
with the associated operators~\cite{filipp-svo-04-qpoly-prl}
$
\textsf{\textbf{F}}(X,Y) \pm \textsf{\textbf{F}}(X,Z) \pm \textsf{\textbf{F}}(Y,Z)
$
with the quantum expectation $\textsf{\textbf{F}}$ yields the much larger quantum bounds
\begin{equation}
-3 <  \textsf{\textbf{F}}(X,Y) \pm \textsf{\textbf{F}}(X,Z) \pm \textsf{\textbf{F}}(Y,Z)   < 3
\label{2020-ex-SZIq}
\end{equation}
that allows a violation of the classical bounds~(\ref{2020-ex-SZI}),
which is a signature of probabilistic contextuality.

\subsection{Functional signatures of contextuality}

There exist configurations of observables that, interpreted classically, serve all kinds of (logical) functions.
(Graph theoretically they are gadgets.)
Usually, they have input and output terminals which, functionally interpreted, serve as arguments and functional values.
Two historic configurations realize either true-implies-false~\cite{kochen2,2018-minimalYIYS}
or true-implies-true functional relations~\cite{kochen1,svozil-2020-hardy}:
if a particular state is preselected on the input terminal then
classical value assignments (implementing omni-existence and noncontextuality)
enforce a particular dependent value assignment---either false or true, respectively---on
the output terminal.

Violations of these dependencies
by quantum probabilities are interpreted as signifying contextuality relative to the assumptions discussed earlier.
In particular, any classical ``hidden variable'' model cannot implement both omni-existence and have noncontextual admissible value assignments~\cite{peres222}.

For the sake of examples we refer to (extensions of)
the Specker bug~\cite{kochen2,2018-minimalYIYS}, or the examples in Refs.~\cite{Ramanathan-18,svozil-2018-whycontexts}
which use finite sets of quantum observables in three-dimensional Hilbert space,
as well as Hardy type configurations~\cite{kochen1,svozil-2020-hardy}
which use finite sets of quantum observables in four- or higher-dimensional~\cite{2018-minimalYIYS} Hilbert space.
All of these gadgets have a classical interpretation in terms of partition logic,
finite automaton models or generalized urn models.
Their respective logico-algebraic structure can be faithfully embedded into extended Boolean algebras;
for instance, $2^n$,
by identification with the union of elements of a partition obtained from analyzing a complete set of $n$ two-valued states~\cite{svozil-2001-eua}.

For the sake of an example we shall review a configuration of 13 observables in seven bi-intertwining contexts (maximal Boolean subalgebras $2^3$) introduced by Kochen and Specker~\cite{kochen2}.
It hypergraph representing contexts as smooth curves (lines) is depicted in Figure~\ref{2021-context-figure-sb}.
\ifx\revtex\undefined
\begin{figure}%[H] %add [H] placement to break table across pages
\else
\begin{figure*}%[H] %add [H] placement to break table across pages
\fi
\centering
\resizebox{0.8\textwidth}{!}{
\begin{tabular}{cccccc}
\begin{tikzpicture}  [scale=0.8]

\tikzstyle{every path}=[line width=1pt]

\newdimen\ms
\ms=0.1cm
\tikzstyle{s1}=[color=red,rectangle,inner sep=3.5]
\tikzstyle{c3}=[circle,inner sep={\ms/8},minimum size=5*\ms]
\tikzstyle{c2}=[circle,inner sep={\ms/8},minimum size=3*\ms]
\tikzstyle{c1}=[circle,inner sep={\ms/8},minimum size=2*\ms]
\tikzstyle{cs1}=[circle,inner sep={\ms/8},minimum size=1*\ms]

% Define positions of all observables

\coordinate (a7) at  (1,2);
\coordinate (a5) at (2,1);
\coordinate (a10) at (0.5,0.5);
\coordinate (a11) at (2,-1);
\coordinate (a1) at (1,-2);
\coordinate (a3) at (0,-1);
\coordinate (a4) at (1.5,0.5);
\coordinate (a9) at (0,1);

% draw contexts

\draw [color=orange] (a1) -- (a11)  coordinate[cs1,fill=white,draw=gray,pos=0.5,label=below right:{\color{black}$a_{12}$}] (a12);
\draw [color=blue] (a11) -- (a9);
\draw [color=red] (a9) -- (a7)  coordinate[cs1,fill=white,draw=gray,pos=0.5,label=above left:{\color{black}$a_8$}] (a8);
\draw [color=green] (a7) -- (a5)  coordinate[cs1,fill=white,draw=gray,pos=0.5,label=above right:{\color{black}$a_6$}] (a6);
\draw [color=gray] (a5) -- (a3);
\draw [color=magenta] (a1) -- (a3)  coordinate[cs1,fill=white,draw=gray,pos=0.5,label= below left:{\color{black}$a_2$}] (a2);
\draw [color=cyan] (a10) -- (a4)  coordinate[cs1,fill=white,draw=gray,pos=0.5,label=above:{\color{black}$a_{13}$}] (a13);

% draw atoms

\draw (a1) coordinate[c2,fill=orange,label=below:$a_1$];
\draw (a1) coordinate[c1,fill=magenta];

\draw (a3) coordinate[c2,fill=magenta,label=left:$a_3$];
\draw (a3) coordinate[c1,fill=gray];

\draw (a11) coordinate[c2,fill=blue,label=right:$a_{11}$];
\draw (a11) coordinate[c1,fill=orange];

\draw (a10) coordinate[c2,fill=cyan,label=below left:$a_{10}$];
\draw (a10) coordinate[c1,fill=blue];

\draw (a5) coordinate[c2,fill=gray,label=right:$a_{5}$];
\draw (a5) coordinate[c1,fill=cyan];

\draw (a9) coordinate[c2,fill=red,label=left:$a_9$];
\draw (a9) coordinate[c1,fill=blue];

\draw (a7) coordinate[c2,fill=gray,label=above:$a_7$];
\draw (a7) coordinate[c1,fill=green];

\draw (a4) coordinate[c2,fill=cyan,label=below right:$a_4$];
\draw (a4) coordinate[c1,fill=gray];

\end{tikzpicture}
&
%%%%%%%%%%%%%%%
&
\begin{tikzpicture}  [scale=0.8]

\tikzstyle{every path}=[line width=1pt]

\newdimen\ms
\ms=0.1cm
\tikzstyle{s1}=[color=red,rectangle,inner sep=3.5]
\tikzstyle{c3}=[circle,inner sep={\ms/8},minimum size=5*\ms]
\tikzstyle{c2}=[circle,inner sep={\ms/8},minimum size=3*\ms]
\tikzstyle{c1}=[circle,inner sep={\ms/8},minimum size=2*\ms]
\tikzstyle{cs1}=[circle,inner sep={\ms/8},minimum size=1*\ms]

% Define positions of all observables

\coordinate (a7) at  (1,2);
\coordinate (a5) at (2,1);
\coordinate (a10) at (0.5,0.5);
\coordinate (a11) at (2,-1);
\coordinate (a1) at (1,-2);
\coordinate (a3) at (0,-1);
\coordinate (a4) at (1.5,0.5);
\coordinate (a9) at (0,1);

% draw contexts

\draw [color=orange] (a1) -- (a11)  coordinate[cs1,fill=white,draw=gray,pos=0.5,label=below right:{\color{black}$p(a_{12})$}] (a12);
\draw [color=blue] (a11) -- (a9);
\draw [color=red] (a9) -- (a7)  coordinate[cs1,fill=white,draw=gray,pos=0.5,label=above left:{\color{black}$p(a_8)$}] (a8);
\draw [color=green] (a7) -- (a5)  coordinate[cs1,fill=white,draw=gray,pos=0.5,label=above right:{\color{black}$p(a_6$)}] (a6);
\draw [color=gray] (a5) -- (a3);
\draw [color=magenta] (a1) -- (a3)  coordinate[cs1,fill=white,draw=gray,pos=0.5,label= below left:{\color{black}$p(a_2)$}] (a2);
\draw [color=cyan] (a10) -- (a4)  coordinate[cs1,fill=white,draw=gray,pos=0.5,label=above:{\color{black}$p(a_{13})$}] (a13);

% draw atoms

\draw (a1) coordinate[c2,fill=orange,label=below:\colorbox{white}{\textcolor{blue}{$p(a_1)=\lambda_1+\lambda_2+\lambda_3$}}];
\draw (a1) coordinate[c1,fill=magenta];

\draw (a3) coordinate[c2,fill=magenta,label=left:$p(a_3)$];
\draw (a3) coordinate[c1,fill=gray];

\draw (a11) coordinate[c2,fill=blue,label=right:$p(a_{11})$];
\draw (a11) coordinate[c1,fill=orange];

\draw (a10) coordinate[c2,fill=cyan,label=below left:$p(a_{10})$];
\draw (a10) coordinate[c1,fill=blue];

\draw (a5) coordinate[c2,fill=gray,label=right:$p(a_{5})$];
\draw (a5) coordinate[c1,fill=cyan];

\draw (a9) coordinate[c2,fill=red,label=left:$p(a_9)$];
\draw (a9) coordinate[c1,fill=blue];

\draw (a7) coordinate[c2,fill=gray,label=above:\colorbox{white}{\textcolor{red}{$p(a_7)=\lambda_6+\lambda_{13}+\lambda_{14}$}}];
\draw (a7) coordinate[c1,fill=green];

\draw (a4) coordinate[c2,fill=cyan,label=below right:$p(a_4)$];
\draw (a4) coordinate[c1,fill=gray];

\end{tikzpicture}
&
%%%%%%%%%%%%%%%%%%%
&
\begin{tikzpicture}  [scale=0.8]

\tikzstyle{every path}=[line width=1pt]

\newdimen\ms
\ms=0.1cm
\tikzstyle{s1}=[color=red,rectangle,inner sep=3.5]
\tikzstyle{c3}=[circle,inner sep={\ms/8},minimum size=5*\ms]
\tikzstyle{c2}=[circle,inner sep={\ms/8},minimum size=3*\ms]
\tikzstyle{c1}=[circle,inner sep={\ms/8},minimum size=2*\ms]
\tikzstyle{cs1}=[circle,inner sep={\ms/8},minimum size=1*\ms]

% Define positions of all observables

\coordinate (a7) at  (1,2);
\coordinate (a5) at (2,1);
\coordinate (a10) at (0.5,0.5);
\coordinate (a11) at (2,-1);
\coordinate (a1) at (1,-2);
\coordinate (a3) at (0,-1);
\coordinate (a4) at (1.5,0.5);
\coordinate (a9) at (0,1);

% draw contexts

\draw [color=orange] (a1) -- (a11)  coordinate[cs1,fill=white,draw=gray,pos=0.5,label=below right:{\scriptsize \color{black}$\begin{pmatrix}0,1,\sqrt{3}\end{pmatrix}$}] (a12);
\draw [color=blue] (a11) -- (a9);
\draw [color=red] (a9) -- (a7)  coordinate[cs1,fill=white,draw=gray,pos=0.5,label=above left:{\scriptsize \color{black}$\begin{pmatrix}-2\sqrt{2},\sqrt{2},-3\sqrt{3}\end{pmatrix}$}] (a8);
\draw [color=green] (a7) -- (a5)  coordinate[cs1,fill=white,draw=gray,pos=0.5,label=above right:{\scriptsize \color{black}$\begin{pmatrix}2\sqrt{2},-1,-3\sqrt{3}\end{pmatrix}$}] (a6);
\draw [color=gray] (a5) -- (a3);
\draw [color=magenta] (a1) -- (a3)  coordinate[cs1,fill=white,draw=gray,pos=0.5,label= below left:{\scriptsize \color{black}$\begin{pmatrix}0, -1, \sqrt{3}]\end{pmatrix}$}] (a2);
\draw [color=cyan] (a10) -- (a4)  coordinate[cs1,fill=white,draw=gray,pos=0.5,label=above:{\tiny \color{black}$\begin{pmatrix}2,-2\sqrt{2},0\end{pmatrix}$}] (a13);

% draw atoms

 \draw (a1) coordinate[c2,fill=orange,label=below:{\colorbox{white}{\textcolor{blue}{$\begin{pmatrix}1,0,0\end{pmatrix}$}}}];
 \draw (a1) coordinate[c1,fill=magenta];

\draw (a3) coordinate[c2,fill=magenta,label=left:{$\begin{pmatrix}0,\sqrt{3},1\end{pmatrix}$}];
\draw (a3) coordinate[c1,fill=gray];

\draw (a11) coordinate[c2,fill=blue,label=right:{$\begin{pmatrix}0,\sqrt{3},-1\end{pmatrix}$}];
\draw (a11) coordinate[c1,fill=orange];

\draw (a10) coordinate[c2,fill=cyan,label=below left:{$\begin{pmatrix}\sqrt{2},1,\sqrt{3}\end{pmatrix}$}];
\draw (a10) coordinate[c1,fill=blue];

\draw (a5) coordinate[c2,fill=gray,label=right:{$\begin{pmatrix}-2\sqrt{2},1,-\sqrt{3}\end{pmatrix}$}];
\draw (a5) coordinate[c1,fill=cyan];

\draw (a9) coordinate[c2,fill=red,label=left:{$\begin{pmatrix}-2\sqrt{2},1,\sqrt{3}\end{pmatrix}$}];
\draw (a9) coordinate[c1,fill=blue];

\draw (a7) coordinate[c2,fill=gray,label=above:{\colorbox{white}{\textcolor{red}{$\begin{pmatrix}1,2\sqrt{2},0\end{pmatrix}$}}}];
\draw (a7) coordinate[c1,fill=green];

\draw (a4) coordinate[c2,fill=cyan,label=below right:{$\begin{pmatrix}\sqrt{2},1,-\sqrt{3}\end{pmatrix}$}];
\draw (a4) coordinate[c1,fill=gray];

\end{tikzpicture}
\\
(a)&&(b)&&(c)\\
\end{tabular}
}
\caption{\label{2021-context-figure-sb} (a) Configuration of 13 observables in seven bi-intertwining contexts serving as a symmetric (with respect to horizontal)
true-implies-false gadget; (b) the associated classical probability distributions obtained by the convex sum
$\lambda_1+\cdots +\lambda_{14}=1$, $\lambda_i\ge 0$, $1 \le i \le 14$;
(c) a quantum representation in terms of a faithful orthogonal vertex labeling
in terms of a vertex representation by vectors, preserving orthogonality of adjacent vertices~\cite{lovasz-79}
of the hypergraph that maximizes
the probability of $a_7$, given $a_1$.}
\ifx\revtex\undefined
\end{figure}
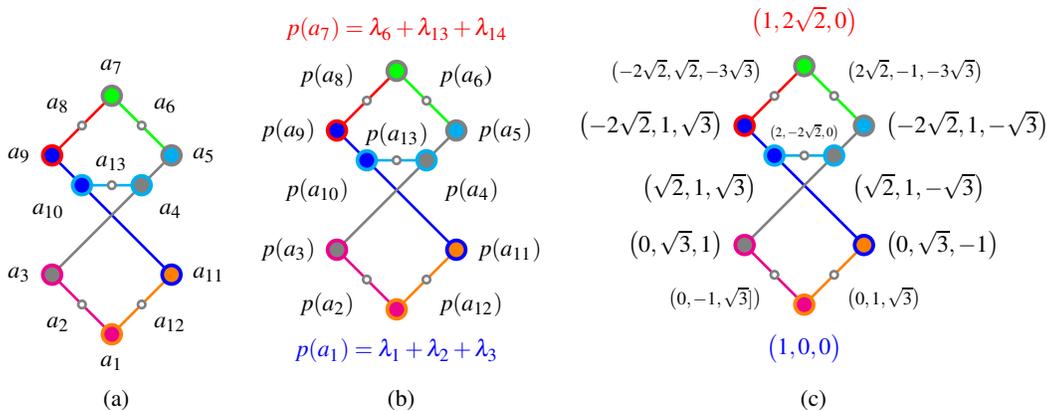%[H] %add [H] placement to break table across pages
\else
\end{figure*}%[H] %add [H] placement to break table across pages
\fi

\begin{table}%[H] %add [H] placement to break table across pages
\centering
%\begin{ruledtabular}
\begin{tabular}{c|cccccccccccccccccccccc}
\toprule
\textbf{\#} & $a_1$ & $a_2$ & $a_3$ & $a_4$ & $a_5$ & $a_6$ & $a_7$ & $a_8$ & $a_9$ & $a_{10}$ & $a_{11}$ & $a_{12}$ & $a_{13}$ \\
\midrule
$v_1$ & \colorbox{white}{\fbox{\textcolor{blue}{$1$}}}&0& 0& 1& 0& 1& 0& 0& 1& 0& 0& 0& 0\\
$v_2$ & \colorbox{white}{\fbox{\textcolor{blue}{$1$}}}&0& 0& 0& 1& 0& 0& 1& 0& 1& 0& 0& 0\\
$v_3$ & \colorbox{white}{\fbox{\textcolor{blue}{$1$}}}&0& 0& 0& 1& 0& 0& 0& 1& 0& 0& 0& 1\\
$v_4$ & 0&1& 0& 1& 0& 1& 0& 1& 0& 0& 1& 0& 0\\
$v_5$ & 0&1& 0& 1& 0& 1& 0& 0& 1& 0& 0& 1& 0\\
$v_6$ & 0&1& 0& 1& 0& 0& \colorbox{white}{\fbox{\textcolor{red}{$1$}}}& 0& 0& 0& 1& 0& 0\\
$v_7$ & 0&1& 0& 0& 1& 0& 0& 1& 0& 1& 0& 1& 0\\
$v_8$ & 0&1& 0& 0& 1& 0& 0& 1& 0& 0& 1& 0& 1\\
$v_9$ & 0&1& 0& 0& 1& 0& 0& 0& 1& 0& 0& 1& 1\\
$v_{10}$ & 0&0& 1& 0& 0& 1& 0& 1& 0& 1& 0& 1& 0\\
$v_{11}$ & 0&0& 1& 0& 0& 1& 0& 1& 0& 0& 1& 0& 1\\
$v_{12}$ & 0&0& 1& 0& 0& 1& 0& 0& 1& 0& 0& 1& 1\\
$v_{13}$ & 0&0& 1& 0& 0& 0& \colorbox{white}{\fbox{\textcolor{red}{$1$}}}& 0& 0& 1& 0& 1& 0\\
$v_{14}$ & 0&0& 1& 0& 0& 0& \colorbox{white}{\fbox{\textcolor{red}{$1$}}}& 0& 0& 0& 1& 0& 1\\
\bottomrule
\end{tabular}
%\end{ruledtabular}
 \caption{\label{2021-context-table-sb}  The 14  $\{0,1\}$-value assignments of the Specker bug configuration. Boxed values indicate nonvanishing contributions of the respective two-valued state to the probabilities of the terminal points $a_1$ and $a_7$ of the gadget.}
 \end{table}

This Specker bug (Specker's ``K\"afer'' configuration can be employed as a classical (noncontextual) true-implies-false (TIF) gadget, as
the two ``terminal points'' $a_1$ and $a_7$ cannot both be ``true'' (value 1) at the same time:
suppose $a_1$ and $a_7$ are both $1$ simultaneously.
Then admissibility demands that $a_3$, $a_5$, $a_9$ as well as $a_{11}$ must all be $0$ simultaneously.
Consequently  $a_4$ and $a_{10}$ must both be $1$ simultaneously---a complete contradiction since admissibility of two-valued $\{0,1\}$-states requires exactly one
element of a context to be 1, all other elements must be 0.
Nevertheless one can be true (have value 1) and the other one false (value 0).
Also they can both be false (value 0): their respective probabilities,
which can be obtained by the convex sum of its 14 two-valued states enumerated in Table~\ref{2021-context-table-sb}
and depicted in in Figure~\ref{2021-context-figure-sb}(b), are
\colorbox{white}{\textcolor{black}{$p(a_1)=\lambda_1+\lambda_2+\lambda_3$}}
and
\colorbox{white}{\textcolor{black}{$p(a_7)=\lambda_6+\lambda_{13}+\lambda_{14}$}}
are mutually exclusive.

Quantization in terms of faithful orthogonal vertex labeling---a
vertex representation by vectors, preserving orthogonality of adjacent vertices~\cite{lovasz-79}---as for instance,
depicted in Figure~\ref{2021-context-figure-sb}(c),
allows the simultaneous preparation by the pre-selection state $a_1$ and the detection of the post-selected state
$a_7$ with nonvanishing
probabilities $\vert     a_7  \cdot a_1  \vert^2 \le \frac{1}{9}$, thereby violating the classical predictions of zero chance that $a_7$ occurs if
$a_1$ occurred.
In this concrete realization with $a_1\equiv \begin{pmatrix}1,0,0\end{pmatrix}^\intercal$ and
$a_7\equiv \frac{1}{3} \begin{pmatrix}1,2\sqrt{2},0\end{pmatrix}^\intercal$
the violation with the classical prediction is maximal~\cite{Belinfante-73,redhead,cabello-1994}.

Also in this case the set of two-valued states allowing a classical co-representation even of complementary observables.
And yet, probabilistic contextuality manifests itself in the functional performance of the respective gadgets at its terminal points.

\subsection{Ad hoc signatures of contextuality}

We just mention without further discussion that there exist other ad hoc methods,
in particular, so-called ``state independent quantum contextuality''~\cite{cabello:210401}, to obtain probabilistic contextuality.
Often the observables of some Kochen-Specker configurations without any two-valued state are taken.
By relaxing the axioms of admissibility mentioned earlier, ``classical'' (relative to these relaxed assumptions) estimates and predictions are obtained
which are violated by experimentally testable quantum predictions~\cite{cabello2020converting}.

\section{Examples of strong contextuality}

In what follows we shall deal with (finite) configurations of observable whose classical interpretations in
terms of its two-valued states is insufficient for an embedding into any Boolean algebra.
Even though there still may exist ``many'' two-valued states associated with classical value assignments,
these assignments may not be able to resolve the structure of quantum observables supporting them.
There exist escalations of the ``smallness'' of the set of two-valued states in terms of
inseparability, nonunitality, or nonexistence that will be briefly reviewed next.
All these instances go beyond classical embeddability (and preservation of the logico-algebraic structure of the associated observables)
as they do not satisfy Kochen and Specker's demarcation criterion~\cite[Theorem~0]{kochen1} for separability.

\subsection{Nonseparability of classical value assignments}

As mentioned there exist finite sets of observables that do not allow separation by classical value assignments.
That is, in such circumstances, no
classical value assignment exists that
is capable to differentiate between, or separating, the individual constituents of some pair of distinct quantum observables.

For the sake of an example of contextuality based on inseparability,
take Kochen and Specker's combo~\cite[Graph~$\Gamma_3$]{kochen1} of intertwining true-implies-true gadgets~\cite[Graph~$\Gamma_1$]{kochen1}
that contains two pairs of observables that cannot be classically separated.
Its hypergraph is depicted in Figure~\ref{2021-context-figure-ksc}.
\ifx\revtex\undefined
\begin{figure}%[H] %add [H] placement to break table across pages
\else
\begin{figure*}%[H] %add [H] placement to break table across pages
\fi
\centering
\resizebox{0.6\textwidth}{!}{
\begin{tikzpicture}  [scale=1]

\tikzstyle{every path}=[line width=1pt]

\newdimen\ms
\ms=0.1cm
\tikzstyle{s1}=[color=red,rectangle,inner sep=3.5]
\tikzstyle{c3}=[circle,inner sep={\ms/8},minimum size=4*\ms]
\tikzstyle{c2}=[circle,inner sep={\ms/8},minimum size=3*\ms]
\tikzstyle{c1}=[circle,inner sep={\ms/8},minimum size=2*\ms]

% Define positions of all observables

% bug #1
\coordinate (a1) at  (1,2);
\coordinate (a2) at (1.5,{(1-(1-0.5)/2)*2});
\coordinate (a3) at (2,1);
\coordinate (a4) at (2,0);
\coordinate (a5) at (2,-1);
\coordinate (a6) at (1.5,{(-0.5-(1-0.5)/2)*2});
\coordinate (a7) at (1,-2);
\coordinate (a8) at (0.5,{(-0.5-(1-0.5)/2)*2});
\coordinate (a9) at (0,-1);
\coordinate (a10) at (0,0);
\coordinate (a11) at (0,1);
\coordinate (a12) at (0.5,{(1-(1-0.5)/2)*2});
\coordinate (a13) at (1,0);

\coordinate (a14) at (4,0);

% bug #2
\coordinate (a21) at  ({6+1},2);
\coordinate (a22) at ({6+1.5},{(1-(1-0.5)/2)*2});
\coordinate (a23) at ({6+2},1);
\coordinate (a24) at ({6+2},0);
\coordinate (a25) at ({6+2},-1);
\coordinate (a26) at ({6+1.5},{(-0.5-(1-0.5)/2)*2});
\coordinate (a27) at ({6+1},-2);
\coordinate (a28) at ({6+0.5},{(-0.5-(1-0.5)/2)*2});
\coordinate (a29) at ({6+0},-1);
\coordinate (a210) at ({6+0},0);
\coordinate (a211) at ({6+0},1);
\coordinate (a212) at ({6+0.5},{(1-(1-0.5)/2)*2});
\coordinate (a213) at ({6+1},0);

% draw contexts

% bug #1
\draw [color=orange] (a1) -- (a3);
\draw [color=blue] (a3) -- (a5);
\draw [color=red] (a5) -- (a7);
\draw [color=green] (a7) -- (a9);
\draw [color=gray] (a9) -- (a11);
\draw [color=magenta] (a11) -- (a1);
\draw [color=cyan] (a10) -- (a4);

\draw [color=lime] (a1) to [out=0,in={90+45}] (a14) to [out={180+90+45},in=180]  (a27);
\draw [color=olive] (a7) to [out=0,in={180+45}] (a14) to [out=45,in=180] (a21);

% bug #2
\draw [color=orange!40!white] (a21) -- (a23);
\draw [color=blue!40!white] (a23) -- (a25);
\draw [color=red!40!white] (a25) -- (a27);
\draw [color=green!40!white] (a27) -- (a29);
\draw [color=gray!40!white] (a29) -- (a211);
\draw [color=magenta!40!white] (a211) -- (a21);
\draw [color=cyan!40!white] (a210) -- (a24);

% draw atoms

% bug 1
\draw (a1) coordinate[c3,fill=lime,label=above:{\colorbox{white}{\textcolor{red}{$a_7$}}}];
\draw (a1) coordinate[c2,fill=orange];
\draw (a1) coordinate[c1,fill=gray];

\draw (a2) coordinate[c1,fill=orange,label=below left:$a_8$];

\draw (a3) coordinate[c2,fill=blue,label=right:$a_9$];
\draw (a3) coordinate[c1,fill=orange];

\draw (a4) coordinate[c2,fill=cyan,label=right:$a_{10}$];
\draw (a4) coordinate[c1,fill=blue];

\draw (a5) coordinate[c2,fill=red,label=right:$a_{11}$];
\draw (a5) coordinate[c1,fill=blue];

\draw (a6) coordinate[c1,fill=red,label=above left:$a_{12}$];

\draw (a7) coordinate[c3,fill=olive,label=below:{\colorbox{white}{\textcolor{blue}{$a_1$}}}];
\draw (a7) coordinate[c2,fill=green];
\draw (a7) coordinate[c1,fill=red];

\draw (a8) coordinate[c1,fill=green,label=below left:$a_2$];

\draw (a9) coordinate[c2,fill=gray,label=left:$a_3$];
\draw (a9) coordinate[c1,fill=green];

\draw (a10) coordinate[c2,fill=gray,label=left:$a_4$];
\draw (a10) coordinate[c1,fill=cyan];

\draw (a11) coordinate[c2,fill=magenta,label=left:$a_5$];
\draw (a11) coordinate[c1,fill=gray];

\draw (a12) coordinate[c1,fill=magenta,label=above left:$a_6$];

\draw (a13) coordinate[c1,fill=cyan,label=above:$a_{13}$];

\draw (a14) coordinate[c2,fill=lime,label=above:{$c$}];
\draw (a14) coordinate[c1,fill=olive];

\draw (a21) coordinate[c3,fill=olive,label=above:{\colorbox{white}{\textcolor{red}{$b_7$}}}];

\draw (a27) coordinate[c3,fill=lime,label=below:{\colorbox{white}{\textcolor{blue}{$b_1$}}}];

% bug 2
\draw (a21) coordinate[c2,fill=orange!40!white];
\draw (a21) coordinate[c1,fill=gray!40!white];

\draw (a22) coordinate[c1,fill=orange!40!white,label=above right:$b_8$];

\draw (a23) coordinate[c2,fill=blue!40!white,label=right:$b_9$];
\draw (a23) coordinate[c1,fill=orange!40!white];

\draw (a24) coordinate[c2,fill=cyan!40!white,label=right:$b_{10}$];
\draw (a24) coordinate[c1,fill=blue!40!white];

\draw (a25) coordinate[c2,fill=red!40!white,label=right:$b_{11}$];
\draw (a25) coordinate[c1,fill=blue!40!white];

\draw (a26) coordinate[c1,fill=red!40!white,label=below right:$b_{12}$];

\draw (a27) coordinate[c2,fill=green!40!white];
\draw (a27) coordinate[c1,fill=red!40!white];

\draw (a28) coordinate[c1,fill=green!40!white,label=above right:$b_2$];

\draw (a29) coordinate[c2,fill=gray!40!white,label=left:$b_3$];
\draw (a29) coordinate[c1,fill=green!40!white];

\draw (a210) coordinate[c2,fill=gray!40!white,label=left:$b_4$];
\draw (a210) coordinate[c1,fill=cyan!40!white];

\draw (a211) coordinate[c2,fill=magenta!40!white,label=left:$b_5$];
\draw (a211) coordinate[c1,fill=gray!40!white];

\draw (a212) coordinate[c1,fill=magenta!40!white,label=below right:$b_6$];

\draw (a213) coordinate[c1,fill=cyan!40!white,label=above:$b_{13}$];

\end{tikzpicture}
}
\caption{\label{2021-context-figure-ksc} The Kochen-Specker ``combo of Specker bugs'' whose set of classical truth assignments formalized by its
two-valued states cannot separate $a_1$ from $b_1$, as well as $a_7$ from $b_7$.}
\ifx\revtex\undefined
\end{figure}
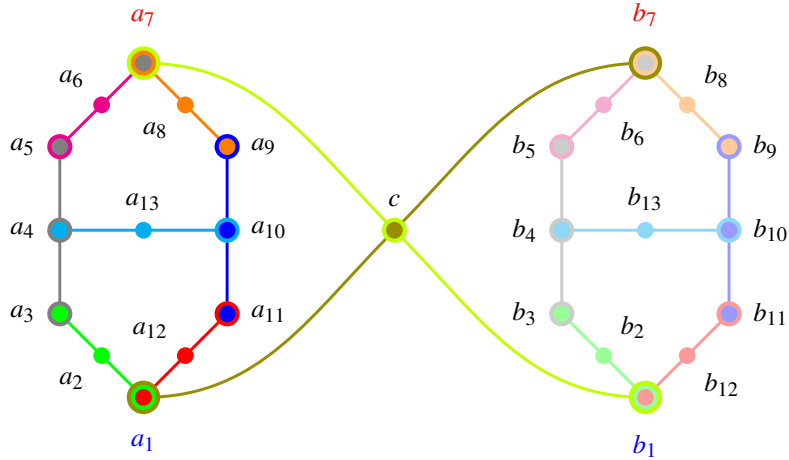%[H] %add [H] placement to break table across pages
\else
\end{figure*}%[H] %add [H] placement to break table across pages
\fi
If $a_1$ is true (has value 1) then $b_1$ has to be true (has value 1), and vice versa.
Likewise,
if $a_7$ is true (has value 1) then $b_7$ has to be true (has value 1), and vice versa.
Therefore,
$a_1$ cannot be classically separated from  $b_1$, and
$a_7$ cannot be classically separated from  $b_7$.
For a proof, note that,  if $a_1$ is assumed to be true, then the true-implies-false Specker bug gadget and
admissibility demands $a_7$ as well as $c$ to be false,
and thus $b_1$ to be true.
Likewise, whenever $b_1$ is  true,
$b_7$ as well as $c$ needs to be false,
and thus $a_1$ to be true. Therefore, $a_1$ cannot be
separated from $b_1$ by any classical means.
A symmetric argument (utilizing the symmetry of the true-implies-false Specker bug gadget)
yields nonseparability of $a_7$  from $b_7$.
The Kochen-Specker combo has no faithful embedding into any ``larger''
Boolean algebra, because any such faithful embedding would allow a classical resolution of the constituents of the pairs
$a_1$ and $b_1$, as well as $a_7$ and $b_7$.

A respective four-dimensional example inspired by Hardy's nonlocal configuration
can be found in Figure~5 of Reference~\cite{svozil-2020-hardy}.
Other explicit experimentally testable cases of inseparability
can be found in a configuration depicted in Figure~2 of Ref.~\cite{tkadlec-96},
as well as in Figure~24.2c analyzed in Table~24.1 of Ref.~\cite{Svozil-2018-p}, based on a configuration introduced in
Figure~2 of Ref.~\cite{2015-AnalyticKS}.

\subsection{Nonunitality of classical value assignments}

Another, even stronger (because it includes and extends inseparability)
form of logical contextuality are collections of observables with a unital set of two-valued states: if interpreted classically such structures
enforce the nonoccurrence (and occurrence) of certain observables.

Two explicit experimentally testable cases of inseparability
are the same as mentioned earlier in a configuration depicted in Figure~2 of Ref.~\cite{tkadlec-96},
as well as in Figure~24.2c analyzed in Table~24.1 of Ref.~\cite{Svozil-2018-p}, based on a configuration introduced in
Figure~2 of Ref.~\cite{2015-AnalyticKS}.

For the sake of demonstration we review this latter configuration depicted in Figure~\ref{2021-context-figure-ACS}.
Analysis of its set of eight two valued states enumerated in Table~\ref{2021-context-table-ACS}  reveals that eight observables, namely
 $a_2$, $a_{13}$, $a_{15}$, $a_{16}$, $a_{17}$, $a_{25}$, $a_{27}$, $a_{36}$ are always 0 (aka false, or nonoccurring) value because they are connected to $a_1$ which has to be 1
(aka true, or always occurring).
As a corollary, those eight observables with simultaneous values 0 cannot be separated from one another with classical means; that is, by two-valued states.

This configuration can, of course, be always subjected to a global rotation, such that its
faithful orthogonal representation,
as enumerated in Table~I of Ref.~\cite{2015-AnalyticKS}, matches $a_1$, or, alternatively, $a_2$.
Thereby, given any pure state, we can as a corollary construct a complete contradiction.
Because given any pure state ${\bf a}$ that can be represented as a unit vector of $\mathbb{R}^3$, we are free to choose
two faithful orthogonal vertex labelings of the hypergraph depicted in Figure~\ref{2021-context-figure-ACS}:
one that matches ${\bf a}$ with $a_1$,
and another one that matches ${\bf a}$ with $a_2$.
Hence ${\bf a}$ would need to be 1 and 0, aka true and false; a complete contradiction.
One could call this argument ``state independent'' because it applies to any pure state ${\bf a}$.

\ifx\revtex\undefined
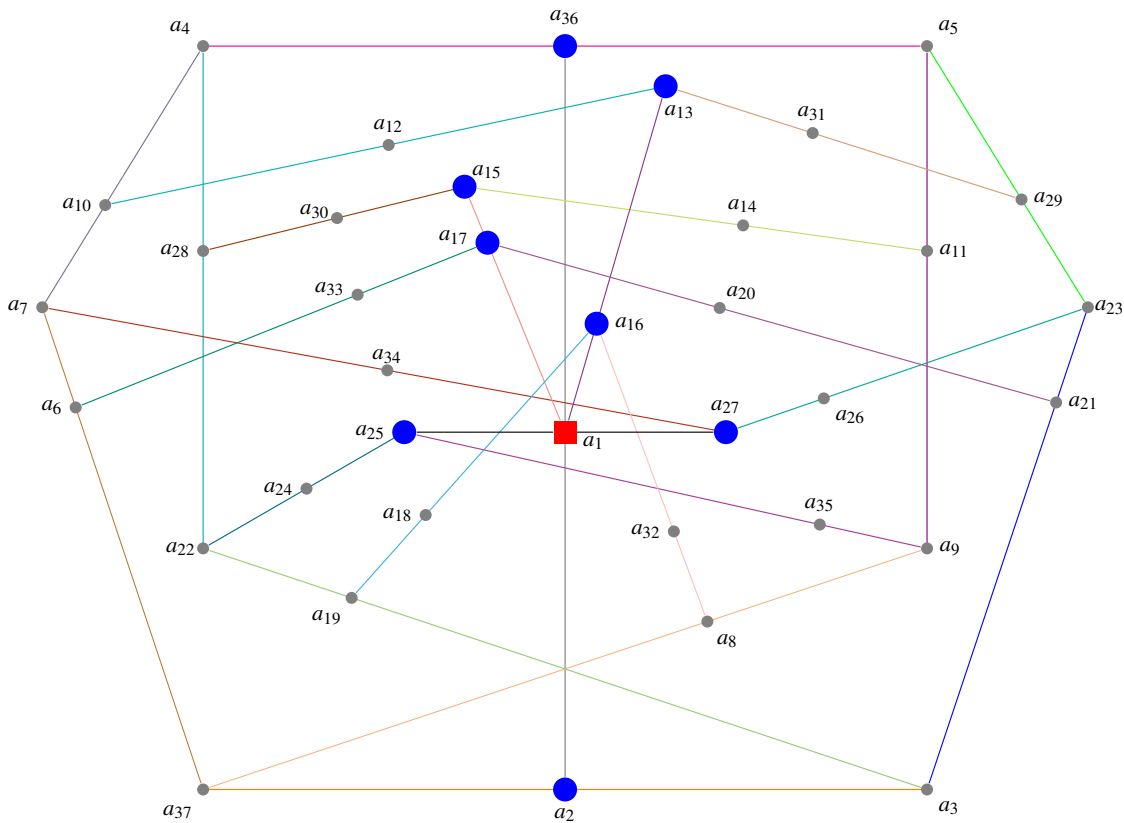
\begin{figure}%[H] %add [H] placement to break table across pages
\else
\begin{figure*}%[H] %add [H] placement to break table across pages
\fi
\centering
\resizebox{0.85\textwidth}{!}{
\begin{tikzpicture}  [scale=0.5, rotate=0]
        %\tikzstyle{every path}=[line width=1.5pt]
%\tikzstyle{c1}=[circle,fill,inner sep=4]
%\tikzstyle{c2}=[circle,fill,inner sep=2.7]
       \tikzstyle{c1}=[color=gray,circle,inner sep=1.5]
       \tikzstyle{c2}=[color=blue,circle,inner sep=3]
        \tikzstyle{s1}=[color=red,rectangle,inner sep=4]
        \tikzstyle{l1}=[draw=none,circle,minimum size=6]

        % Define positions of all observables

\draw [color=orange]  (4,0)  coordinate[c1,fill,label=260:{\color{black}\footnotesize $a_{37}$}] (b) -- (13,0)    coordinate[c2,fill,label={[label distance=-1]270:{\footnotesize \color{black}  $a_2$}}] (2) -- (22,0)  coordinate[c1,fill,label={[label distance=-1]315:{\footnotesize \color{black}  $a_3$}}] (3);
\draw [color=blue] (3) -- (26,12)  coordinate[c1,fill,pos=0.8,label={[label distance=-1]0:{\footnotesize \color{black}  $a_{21}$}}] (21) coordinate[c1,fill,label={[label distance=-3]0:{\footnotesize \color{black}  $a_{23}$}}] (23);
\draw [color=green] (23) -- (22,18.5) coordinate[c1,fill,pos=0.4,label={[label distance=-1]0:{\footnotesize \color{black}  $a_{29}$}}] (29) coordinate[c1,fill,label={[label distance=-1]45:{\footnotesize \color{black}  $a_5$}}] (5);
\draw [color=magenta] (5)-- (13,18.5)coordinate[c2,fill,label=90:{\color{black}\footnotesize $a_{36}$}] (a) -- (4,18.5)  coordinate[c1,fill,label={[label distance=-1]135:{\footnotesize \color{black}  $a_4$}}] (4);
\draw [color=CadetBlue] (4) -- (0,12)   coordinate[c1,fill,pos=0.6,label={[label distance=-1]180:{\footnotesize \color{black}  $a_{10}$}}] (10)  coordinate[c1,fill,label={[label distance=-1]180:{\footnotesize \color{black}  $a_7$}}] (7);
\draw [color=brown] (7) -- (b)      coordinate[c1,fill,pos=0.2,label={[label distance=-1]180:{\footnotesize \color{black}  $a_6$}}] (6);

        \draw [color=gray] (a) -- (2) coordinate[s1,fill,pos=0.52,label={[label distance=-1, yshift=2]357.5:{\footnotesize \color{black}  $a_1$}}] (1);

        \draw [color=violet] (5) -- (22,6) coordinate[c1,fill,pos=0.4,label={[label distance=-1]0:{\footnotesize \color{black}  $a_{11}$}}] (11) coordinate[c1,fill,label={[label distance=-1]0:{\footnotesize \color{black}  $a_9$}}] (9);

\draw [color=Apricot] (9) -- (b) coordinate[c1,fill,pos=0.3,label={[label distance=-1]280:{\footnotesize \color{black}  $a_8$}}] (8);

\draw [color=TealBlue] (4) -- (4,6) coordinate[c1,fill,pos=0.4,label={[label distance=-1]180:{\footnotesize \color{black}  $a_{28}$}}] (28) coordinate[c1,fill,label={[label distance=-3]180:{\footnotesize \color{black}  $a_{22}$}}] (22);
\draw [color=YellowGreen] (22) -- (3) coordinate[c1,fill,pos=0.2,label={[label distance=-1]260:{\footnotesize \color{black}  $a_{19}$}}] (19);

        \coordinate (25) at ([xshift=-4cm]1);
        \coordinate (27) at ([xshift=4cm]1);

\draw [color=MidnightBlue]  (22) -- (25) coordinate[c1,fill,pos=0.5,label={[label distance=-1]180:{\footnotesize \color{black}  $a_{24}$}}] (24) coordinate[c2,fill,label={[label distance=-1]180:{\footnotesize \color{black}  $a_{25}$}}] (25);
\draw [color=Mulberry] (25) -- (9) coordinate[c1,fill,pos=0.8,label={[label distance=-1]90:{\footnotesize \color{black}  $a_{35}$}}] (35);

\draw [color=BrickRed]  (7) -- (27) coordinate[c1,fill,pos=0.5,label={[label distance=-3]90:{\footnotesize \color{black}  $a_{34}$}}] (34) coordinate[c2,fill,label={[label distance=-1]90:{\footnotesize \color{black}  $a_{27}$}}] (27);
\draw [color=Emerald] (27) -- (23) coordinate[c1,fill,pos=0.25,label={[label distance=-1]320:{\footnotesize \color{black}  $a_{26}$}}] (26);

\draw [color=BlueGreen]  (10) -- (15.5,17.5) coordinate[c1,fill,pos=0.5,label={[label distance=-1]90:{\footnotesize \color{black}  $a_{12}$}}] (12) coordinate[c2,fill,label={[label distance=-1,xshift=5]270:{\footnotesize \color{black}  $a_{13}$}}] (13);
\draw [color=Tan] (13) -- (29) coordinate[c1,fill,pos=0.4,label={[label distance=-1]90:{\footnotesize \color{black}  $a_{31}$}}] (31);

\draw [color=RawSienna]  (28) -- (10.5,15) coordinate[c1,fill,pos=0.5,label={[label distance=-3, yshift=-3]160:{\footnotesize \color{black}  $a_{30}$}}] (30) coordinate[c2,fill,label={[label distance=-5]45:{\footnotesize \color{black}  $a_{15}$}}] (15);
\draw [color=SpringGreen] (15) -- (11) coordinate[c1,fill,pos=0.6,label={[label distance=-1]90:{\footnotesize \color{black}  $a_{14}$}}] (14);

\draw [color=Salmon]  (15) -- (1) coordinate[c2,fill,pos=0.2,label={[label distance=-1, yshift=2]180:{\footnotesize \color{black}  $a_{17}$}}] (17);
\draw [color=Fuchsia] (1)-- (13) coordinate[c2,fill,pos=0.3,label={[label distance=-1]0:{\footnotesize \color{black}  $a_{16}$}}] (16);

\draw [color=CornflowerBlue]  (19) -- (16) coordinate[c1,fill,pos=0.3,label={[label distance=-1]180:{\footnotesize \color{black}  $a_{18}$}}] (18);
\draw [color=pink] (16) -- (8) coordinate[c1,fill,pos=0.7,label={[label distance=-1]180:{\footnotesize \color{black}  $a_{32}$}}] (32);

\draw [color=PineGreen]  (6) -- (17) coordinate[c1,fill,pos=0.7,label={[label distance=-1, yshift=2]180:{\footnotesize \color{black}  $a_{33}$}}] (33);
\draw [color=DarkOrchid] (17) -- (21) coordinate[c1,fill,pos=0.4,label={[label distance=-3]20:{\footnotesize \color{black}  $a_{20}$}}] (20);

\draw [color=black] (25) -- (1) -- (27);

\end{tikzpicture}
}
\caption{\label{2021-context-figure-ACS} A configuration of quantum observables with a nonunital set of classical two-valued states
in three-dimensional Hilbert space~\cite{Svozil-2018-p}. Admissibility demands that proposition $a_1$ must be true (value 1);
and the adjacent propositions $a_2$, $a_{13}$, $a_{15}$, $a_{16}$, $a_{17}$, $a_{25}$, $a_{27}$, $a_{36}$
sharing hyperedges with $a_1$ must be false (value 0).
All other observables are either 0 or 1, depending on the respective two-valued state enumerated in Table~\ref{2021-context-table-ACS}.}
\ifx\revtex\undefined
\end{figure}%[H] %add [H] placement to break table across pages
\else
\end{figure*}%[H] %add [H] placement to break table across pages
\fi

\ifx\revtex\undefined
\begin{table}%[H] %add [H] placement to break table across pages
\else
\begin{table*}%[H] %add [H] placement to break table across pages
\fi
\centering
\setcounter{MaxMatrixCols}{40}
\resizebox{1\textwidth}{!}{
\begin{tabular}{c|cccccccccccccccccccccccccccccccccccccccccccccccccccccccccc}
\toprule
\# & $a_{ 1}$ & $a_{ 2}$& $a_{ 3}$& $a_{ 4}$& $a_{ 5}$& $a_{ 6}$& $a_{ 7}$& $a_{ 8}$& $a_{ 9}$& $a_{10}$& $a_{11}$& $a_{12}$& $a_{13}$& $a_{14}$& $a_{15}$& $a_{16}$& $a_{17}$& $a_{18}$&$a_{19}$\\
\midrule
1&\colorbox{white}{\fbox{\textcolor{red}{1}}}&\colorbox{white}{\fbox{\textcolor{blue}{0}}}& 0& 1& 0& 0& 0& 0& 0& 0& 1& 1&\colorbox{white}{\fbox{\textcolor{blue}{0}}}& 0&\colorbox{white}{\fbox{\textcolor{blue}{0}}}&\colorbox{white}{\fbox{\textcolor{blue}{0}}}&\colorbox{white}{\fbox{\textcolor{blue}{0}}}& 0& 1\\%& 0& 1& 0& 0& 1&\colorbox{white}{\fbox{\textcolor{blue}{0}}}& 1&\colorbox{white}{\fbox{\textcolor{blue}{0}}}& 0& 1& 1& 0& 1& 1& 1& 1&\colorbox{white}{\fbox{\textcolor{blue}{0}}}& 1&\\
2&\colorbox{white}{\fbox{\textcolor{red}{1}}}&\colorbox{white}{\fbox{\textcolor{blue}{0}}}& 0& 1& 0& 0& 0& 0& 0& 0& 1& 1&\colorbox{white}{\fbox{\textcolor{blue}{0}}}& 0&\colorbox{white}{\fbox{\textcolor{blue}{0}}}&\colorbox{white}{\fbox{\textcolor{blue}{0}}}&\colorbox{white}{\fbox{\textcolor{blue}{0}}}& 0& 1\\%& 1& 0& 0& 1& 1&\colorbox{white}{\fbox{\textcolor{blue}{0}}}& 0&\colorbox{white}{\fbox{\textcolor{blue}{0}}}& 0& 0& 1& 1& 1& 1& 1& 1&\colorbox{white}{\fbox{\textcolor{blue}{0}}}& 1&\\
3&\colorbox{white}{\fbox{\textcolor{red}{1}}}&\colorbox{white}{\fbox{\textcolor{blue}{0}}}& 0& 0& 1& 0& 0& 0& 0& 1& 0& 0&\colorbox{white}{\fbox{\textcolor{blue}{0}}}& 1&\colorbox{white}{\fbox{\textcolor{blue}{0}}}&\colorbox{white}{\fbox{\textcolor{blue}{0}}}&\colorbox{white}{\fbox{\textcolor{blue}{0}}}& 0& 1\\%& 0& 1& 0& 0& 1&\colorbox{white}{\fbox{\textcolor{blue}{0}}}& 1&\colorbox{white}{\fbox{\textcolor{blue}{0}}}& 1& 0& 0& 1& 1& 1& 1& 1&\colorbox{white}{\fbox{\textcolor{blue}{0}}}& 1&\\
4&\colorbox{white}{\fbox{\textcolor{red}{1}}}&\colorbox{white}{\fbox{\textcolor{blue}{0}}}& 0& 0& 1& 0& 0& 0& 0& 1& 0& 0&\colorbox{white}{\fbox{\textcolor{blue}{0}}}& 1&\colorbox{white}{\fbox{\textcolor{blue}{0}}}&\colorbox{white}{\fbox{\textcolor{blue}{0}}}&\colorbox{white}{\fbox{\textcolor{blue}{0}}}& 1& 0\\%& 0& 1& 1& 0& 0&\colorbox{white}{\fbox{\textcolor{blue}{0}}}& 1&\colorbox{white}{\fbox{\textcolor{blue}{0}}}& 0& 0& 1& 1& 1& 1& 1& 1&\colorbox{white}{\fbox{\textcolor{blue}{0}}}& 1&\\
5&\colorbox{white}{\fbox{\textcolor{red}{1}}}&\colorbox{white}{\fbox{\textcolor{blue}{0}}}& 1& 1& 0& 1& 0& 1& 0& 0& 1& 1&\colorbox{white}{\fbox{\textcolor{blue}{0}}}& 0&\colorbox{white}{\fbox{\textcolor{blue}{0}}}&\colorbox{white}{\fbox{\textcolor{blue}{0}}}&\colorbox{white}{\fbox{\textcolor{blue}{0}}}& 1& 0\\%& 1& 0& 0& 0& 1&\colorbox{white}{\fbox{\textcolor{blue}{0}}}& 1&\colorbox{white}{\fbox{\textcolor{blue}{0}}}& 0& 1& 1& 0& 0& 0& 1& 1&\colorbox{white}{\fbox{\textcolor{blue}{0}}}& 0&\\
6&\colorbox{white}{\fbox{\textcolor{red}{1}}}&\colorbox{white}{\fbox{\textcolor{blue}{0}}}& 1& 1& 0& 1& 0& 0& 1& 0& 0& 1&\colorbox{white}{\fbox{\textcolor{blue}{0}}}& 1&\colorbox{white}{\fbox{\textcolor{blue}{0}}}&\colorbox{white}{\fbox{\textcolor{blue}{0}}}&\colorbox{white}{\fbox{\textcolor{blue}{0}}}& 1& 0\\%& 1& 0& 0& 0& 1&\colorbox{white}{\fbox{\textcolor{blue}{0}}}& 1&\colorbox{white}{\fbox{\textcolor{blue}{0}}}& 0& 1& 1& 0& 1& 0& 1& 0&\colorbox{white}{\fbox{\textcolor{blue}{0}}}& 0&\\
7&\colorbox{white}{\fbox{\textcolor{red}{1}}}&\colorbox{white}{\fbox{\textcolor{blue}{0}}}& 1& 0& 1& 1& 0& 1& 0& 1& 0& 0&\colorbox{white}{\fbox{\textcolor{blue}{0}}}& 1&\colorbox{white}{\fbox{\textcolor{blue}{0}}}&\colorbox{white}{\fbox{\textcolor{blue}{0}}}&\colorbox{white}{\fbox{\textcolor{blue}{0}}}& 1& 0\\%& 1& 0& 0& 0& 1&\colorbox{white}{\fbox{\textcolor{blue}{0}}}& 1&\colorbox{white}{\fbox{\textcolor{blue}{0}}}& 1& 0& 0& 1& 0& 0& 1& 1&\colorbox{white}{\fbox{\textcolor{blue}{0}}}& 0&\\
8&\colorbox{white}{\fbox{\textcolor{red}{1}}}&\colorbox{white}{\fbox{\textcolor{blue}{0}}}& 1& 0& 1& 0& 1& 1& 0& 0& 0& 1&\colorbox{white}{\fbox{\textcolor{blue}{0}}}& 1&\colorbox{white}{\fbox{\textcolor{blue}{0}}}&\colorbox{white}{\fbox{\textcolor{blue}{0}}}&\colorbox{white}{\fbox{\textcolor{blue}{0}}}& 1& 0\\%& 1& 0& 0& 0& 1&\colorbox{white}{\fbox{\textcolor{blue}{0}}}& 1&\colorbox{white}{\fbox{\textcolor{blue}{0}}}& 1& 0& 0& 1& 0& 1& 0& 1&\colorbox{white}{\fbox{\textcolor{blue}{0}}}& 0&\\
\midrule
\# &  $a_{20}$& $a_{21}$ & $a_{22}$&$a_{23}$& $a_{24}$& $a_{25}$& $a_{26}$& $a_{27}$& $a_{28}$& $a_{29}$& $a_{30}$& $a_{31}$& $a_{32}$& $a_{33}$& $a_{34}$& $a_{35}$& $a_{36}$ & $a_{37}$\\
\midrule
1&  0& 1& 0& 0& 1&\colorbox{white}{\fbox{\textcolor{blue}{0}}}& 1&\colorbox{white}{\fbox{\textcolor{blue}{0}}}& 0& 1& 1& 0& 1& 1& 1& 1&\colorbox{white}{\fbox{\textcolor{blue}{0}}}& 1&\\
2&  1& 0& 0& 1& 1&\colorbox{white}{\fbox{\textcolor{blue}{0}}}& 0&\colorbox{white}{\fbox{\textcolor{blue}{0}}}& 0& 0& 1& 1& 1& 1& 1& 1&\colorbox{white}{\fbox{\textcolor{blue}{0}}}& 1&\\
3&  0& 1& 0& 0& 1&\colorbox{white}{\fbox{\textcolor{blue}{0}}}& 1&\colorbox{white}{\fbox{\textcolor{blue}{0}}}& 1& 0& 0& 1& 1& 1& 1& 1&\colorbox{white}{\fbox{\textcolor{blue}{0}}}& 1&\\
4&  0& 1& 1& 0& 0&\colorbox{white}{\fbox{\textcolor{blue}{0}}}& 1&\colorbox{white}{\fbox{\textcolor{blue}{0}}}& 0& 0& 1& 1& 1& 1& 1& 1&\colorbox{white}{\fbox{\textcolor{blue}{0}}}& 1&\\
5&  1& 0& 0& 0& 1&\colorbox{white}{\fbox{\textcolor{blue}{0}}}& 1&\colorbox{white}{\fbox{\textcolor{blue}{0}}}& 0& 1& 1& 0& 0& 0& 1& 1&\colorbox{white}{\fbox{\textcolor{blue}{0}}}& 0&\\
6&  1& 0& 0& 0& 1&\colorbox{white}{\fbox{\textcolor{blue}{0}}}& 1&\colorbox{white}{\fbox{\textcolor{blue}{0}}}& 0& 1& 1& 0& 1& 0& 1& 0&\colorbox{white}{\fbox{\textcolor{blue}{0}}}& 0&\\
7&  1& 0& 0& 0& 1&\colorbox{white}{\fbox{\textcolor{blue}{0}}}& 1&\colorbox{white}{\fbox{\textcolor{blue}{0}}}& 1& 0& 0& 1& 0& 0& 1& 1&\colorbox{white}{\fbox{\textcolor{blue}{0}}}& 0&\\
8&  1& 0& 0& 0& 1&\colorbox{white}{\fbox{\textcolor{blue}{0}}}& 1&\colorbox{white}{\fbox{\textcolor{blue}{0}}}& 1& 0& 0& 1& 0& 1& 0& 1&\colorbox{white}{\fbox{\textcolor{blue}{0}}}& 0&\\
\midrule
\bottomrule
\end{tabular}
}
\caption{\label{2021-context-table-ACS} The eight two-valued states on the configuration depicted in Figure~\ref{2021-context-figure-ACS}. Boxes indicate fixed values.}
\ifx\revtex\undefined
\end{table}%[H] %add [H] placement to break table across pages
\else
\end{table*}%[H] %add [H] placement to break table across pages
\fi

\subsection{Nonexistence of classical value assignments}

The most extreme form of strong contextuality occurs if the respective structure of observables allows no classical interpretation whatsoever.
This result had already been announced by Specker in 1960~\cite{specker-60},
and is nowadays called the Kochen-Specker theorem~\cite[Graph~$\Gamma_2$]{kochen1}.
It has been perceived~\cite{ZirlSchl-65,kamber65} as a direct consequence of Gleason's theorem~\cite{Gleason}.

One may, in a certain sense and relative to the mathematical means employed,
extend these results by proving that there exist finite configurations of observables
that do not allow any classical value definite existence beyond a single classical value assignment,
and the (continuity of) contexts containing this extreme case.
Proofs relative to global and total classical value assignments are in Refs.~\cite{pitowsky:218,hru-pit-2003}.
Similar results are obtained with weaker assumptions allowing partial value assignments in
Refs.~\cite{2012-incomput-proofsCJ,PhysRevA.89.032109,2015-AnalyticKS}

\section{Further observations}

%Let me add some pertinent issues:
%First, I would like to raise some concerns about
%a method of ``collapsing'' probabilities, also called the ``support'' of a standard probabilistic Hardy model~\cite{Abramsky2011},
%such that any non-zero probability is mapped onto one,
%which serves as the basis of various ``logic'' arguments%~\cite{Abramsky2012,Abramsky2015,Abramsky2014,Abramsky-2017,Abramsky2018}.
%This ``logical'' method works well for particular gadgets such as true-implies-false gadgets because
%they are symmetrical with respect to their ``terminals'': true-implies-false gadgets
%also work ``the other way round'' by inverting the terminal observables,
%but this is no longer the case for true-implies-true gadgets gadgets.
%In general by effectively ``collapsing'' all two-valued states, one might lose important relational information.
%
%Second, to pronounce that contextuality ``supplies the magic for quantum computation''
%suggests that contextuality might be some computational capacity or resource that raises high expectations
%based on a rather vague notion. One may even argue that contextuality might indicate restrictions rather than extensions
%of classical computations.

\subsection{Omni-existence}

The use of the term ``contextual'' might imply or implicitly suggest, without direct empirical evidence,
a form of omni-existence. But omni-existence is a metaphysical concept because it lacks any direct operational test. Those arguments involve counterfactuals~\cite{specker-60}.

A generalized Jaynes' principle is called ``plausible reasoning'':
one should not introduce unnecessary epistemic bias, superficial information, and individual ontologic projections into empirical evidence
but rather stick to the ``knowable'' facts.
In Jaynes' words~\cite[Section~10.11, p.~331]{jaynes}, {``the onus is always on the user $\ldots$~that the full extent of his ignorance is also
properly represented''}.

Therefore, it might be more appropriate to talk about ``quantum indeterminacy'' as Pitowsky did~\cite{pitowsky:218,hru-pit-2003},
and to allow partial functions and value indefiniteness.
Partial functions have been first conceptualized~\cite{Kleene1936} in theoretical computer science
to cope with and formalize computability; in particular, with the recursive unsolvability of the halting problem.
They are essential in the theory of recursive functions
and indicate lack of capacities that go beyond certain limits of consistent formal expressibility.
It thus might me more appropriate to use the terms ``partial functions'' and ``value indefinite''
instead of ``contextuality''~\cite{2012-incomput-proofsCJ,Abbott:2010uq,PhysRevA.89.032109,2015-AnalyticKS}.

We conjecture that because of Pitowsky's principle of indeterminism~\cite{pitowsky:218,hru-pit-2003}
and newer theorems allowing partial functions as value assignments~\cite{2015-AnalyticKS},
the ``message'' of the quantum is straightforward: quantum systems are defined in their frame of preparation, and undefined in directions
other than perpendicular or collinear.
%Insistence in the assumption of omni-definiteness yields to ``contextual'' as well as stochastic value assignments.
%The respective measurement outcomes do not directly reflect any objective property prior to measurement,
%but on the complexion of the entire entangled state including the measurement apparatus.

\subsection{Is contextuality haunted?}

It should be kept in mind that there exists only indirect empirical tests of contextuality invoking counterfactuals.
Indeed, any experimentally verifiable contextuality remains indirect (or ``haunted'') and not direct, as
quantum mechanics predicts the absence of
direct verifications~\cite{svozil-1999-haunted-qc,svozil:040102,Griffiths2017,Griffiths2019}:
if say, two contexts $\{a,b,c\}$ and $\{a,d,e\}$ intertwining at observable $a$ are considered,
quantum mechanics is not ambiguous about the outcome corresponding to $a$, regardless of the context measured.
This can be directly experimentally tested on a single quantized particle, or on entangled particle pairs.
A remaining ``haunted'' context-dependence might manifest itself in a hidden and uncontrollable outcome dependence of the remaining
complementary observable pairs $\{b,c\}$ and $\{d,e\}$.
In four or higher dimensions this applies also to all quantum observables common to different contexts.
For instance, for a configuration $\{a,b,c,d\}$ and $\{a,b,e,f\}$,
quantum mechanics predicts that the observables $a$ and $b$ are noncontextual, whereas $\{c,d\}$ and $\{e,f\}$
might show (hidden and haunted) outcome dependence.

\subsection{Historic aspects regarding the importance of embeddability for Specker}

Let me add some afterthoughts on Kochen and Specker's demarcation criterion~\cite[Theorem~0]{kochen1}
for structure-preserving (non)embeddability mentioned earlier
as a decisive benchmark or measure for strong contextuality or value indefiniteness.
I consider it not entirely unreasonable to speculate that Specker meant the lack of embeddability by announcing~\cite{specker-60}:
{``An elementary geometrical argument shows that such an assignment is
impossible and that therefore (aside from the exceptions noted above) no
consistent prediction concerning a quantum mechanical system is possible.''}
Indeed, the `Comment' section of the respective article in Specker's `Selecta'~\cite[p.~385]{specker-ges}
explicitly notes (the references are updated to match the current ones used here):
{``The impossibility to embed the lattice of subspaces of $\mathbb{R}^3$
into a Boolean algebra, mentioned at the end of~\cite{specker-60}, is proved in~\cite{kochen1}
(theorem 1 and subsequent remarks).''}

\section{Summary}

Numerous notions of contextuality exist in the literature on quantum foundations.
The concept of ``contextuality'' is often used in ambiguous or contradictory ways.
This paper has taken a logic-based approach to clarify, identify, and categorize these different notions.
In particular, we propose to categorize different notions of contextuality into two major groups:
probabilistic and strong notions of contextuality.
We suggest using Kochen and Specker's demarcation Theorem~0~\cite{kochen1} as a criterion to differentiate between those groups.

The following review summarises and concentrates the issues raised.
\begin{itemize}
\item[(i)]
The supposition of a well-defined physical operationalization of the properties associated with quantum observables, and,
in particular, omni-existence lies at the basis of current so-called empirical tests of contextuality.
This does not take into account the entanglement between the object under observation and the measurement apparatus.
However, such a conception of measurement entails that the constituents of the entangled object-apparatus state are in no definite individual state.

\item[(ii)]
As long as the explicit functional context-dependence of quantum observables common to different contexts is directly tested
it is absent in quantized systems. Therefore, it might be ``haunted',' as such dependence may only occur indirectly, and
without direct experimental testability.

\item[(iii)]
It might be prudent to differentiate between the logico-algebraic structure formed by the observables via
complementarity and the probability distributions such logics can or cannot support.

\item[(iv)]
Kochen and Specker gave a ``demarcation criterion''
for nonembeddability in terms of (in)separability:
if the set of two-valued states on the logic can discriminate between every pair of atomic observables
(aka elementary propositions in the sense of Birkhoff and von Neumann)
then it can support classical models (and also quantum ones if there exist faithful orthogonal representations).
If no classical value assignment aka two-valued state can separate between two observables, then
embeddability in some presumably extended Boolean algebra breaks down.
Such situations can be termed strong contextuality.

\item[(v)]
Current empirical corroborations of contextuality associated with Boole-Bell-type inequalities,
as long as they are based on hull computations of classical value assignments on suitable ensembles of quantum observables
with separating sets of two-valued states encoding these value assignments,
are merely about probabilistic contextuality.
The same is true for separable ensembles of quantum observables with functional relations on endpoints.

%\item[(vi)]
%In the spirit of theoretical computer science and of Itamar Pitowsky I suggest~\cite{Svozil-2018-p} to call these properties ``partial value assignments'',
%``value indefinite', or ``indeterminate''.

\item[(vi)]
Strong contextuality implies probabilistic contextuality because the former always indicates
some ``essential'' scarcity of two-valued states associated with classical truth assignments.
Because, by convex summation, two-valued states form the basis for classical probability distributions,
any ``essential'' lack thereof indicates limits to classical physical phenomenology that provide distinctions from quantized systems.
Essential here stands for nonseparating, nonunital, or in its strongest form not existing.
\end{itemize}

These matters are pertinent not only to foundational questions but also to the computational capacities of quantized systems.

\ifx\revtex\undefined

\section*{Acknowledgments}

This research was funded in whole, or in part, by the Austrian Science Fund (FWF), Project No. I 4579-N. For the purpose of open access, the author has applied a CC BY public copyright licence to any Author Accepted Manuscript version arising from this submission.

%The author acknowledge TU Wien Bibliothek for financial support through its Open Access Funding Programme.

%I kindly thank the IFISC (Institute for Cross-Disciplinary Physics and Complex Systems),
%a joint research institute of the University of the Balearic Islands (UIB) and the Spanish National Research Council (CSIC)
%for hospitality during my stay in February 2021;
%and Adan Cabello as well as Jose Portillo of the University of Sevilla for our ongoing cooperation and during this stay.

%I kindly acknowledge discussions with and suggestions by
%Cristian Calude, Kelly James Clark, Silvia Jonas, Jeffrey Koperski, Irem Kurtsal, Emil Salim, Mohammad Hadi Shekarriz, and Noson S. Yanofsky.

I thank Ad\'an Cabello for sharing his thoughts and questions on the time of origin of what is nowadays called~\cite{Stigler1980} Kochen-Specker theorem.

All misconceptions and errors are mine.

The author declares no conflict of interest.

\section*{References}

%\bibliography{svozil}

%%%%%%%%%%%%%%%%%%%%%%%%%%%%%%%%%%%%%%%%%%%%%%%%%%%%%%%%%%%

%%%%%%%%%%%%%%%%%%%%%%%%%%%%%%%%%%%%%%%%%%%%%%%%%%%%%%%%%%%

\else

\begin{acknowledgments}

This research was funded in whole, or in part, by the Austrian Science Fund (FWF), Project No. I 4579-N. For the purpose of open access, the author has applied a CC BY public copyright licence to any Author Accepted Manuscript version arising from this submission.

The authors acknowledge TU Wien Bibliothek for financial support through its Open Access Funding Programme.

%I kindly thank the IFISC (Institute for Cross-Disciplinary Physics and Complex Systems),
%a joint research institute of the University of the Balearic Islands (UIB) and the Spanish National Research Council (CSIC)
%for hospitality during my stay in February 2021;
%and Adan Cabello as well as Jose Portillo of the University of Sevilla for our ongoing cooperation and during this stay.

%I kindly acknowledge discussions with and suggestions by
%Cristian Calude, Kelly James Clark, Silvia Jonas, Jeffrey Koperski, Irem Kurtsal, Emil Salim, Mohammad Hadi Shekarriz, and Noson S. Yanofsky.

I thank Ad\'an Cabello for sharing his thoughts and questions on the time of origin of what is nowadays called~\cite{Stigler1980} Kochen-Specker theorem.

All misconceptions and errors are mine.

The author declares no conflict of interest.
\end{acknowledgments}

%\bibliography{svozil}

\begin{thebibliography}{79}%
\makeatletter
\providecommand \@ifxundefined [1]{%
 \@ifx{#1\undefined}
}%
\providecommand \@ifnum [1]{%
 \ifnum #1\expandafter \@firstoftwo
 \else \expandafter \@secondoftwo
 \fi
}%
\providecommand \@ifx [1]{%
 \ifx #1\expandafter \@firstoftwo
 \else \expandafter \@secondoftwo
 \fi
}%
\providecommand \natexlab [1]{#1}%
\providecommand \enquote  [1]{``#1''}%
\providecommand \bibnamefont  [1]{#1}%
\providecommand \bibfnamefont [1]{#1}%
\providecommand \citenamefont [1]{#1}%
\providecommand \href@noop [0]{\@secondoftwo}%
\providecommand \href [0]{\begingroup \@sanitize@url \@href}%
\providecommand \@href[1]{\@@startlink{#1}\@@href}%
\providecommand \@@href[1]{\endgroup#1\@@endlink}%
\providecommand \@sanitize@url [0]{\catcode `\\12\catcode `\$12\catcode
  `\&12\catcode `\#12\catcode `\^12\catcode `\_12\catcode `\%12\relax}%
\providecommand \@@startlink[1]{}%
\providecommand \@@endlink[0]{}%
\providecommand \url  [0]{\begingroup\@sanitize@url \@url }%
\providecommand \@url [1]{\endgroup\@href {#1}{\urlprefix }}%
\providecommand \urlprefix  [0]{URL }%
\providecommand \Eprint [0]{\href }%
\providecommand \doibase [0]{https://doi.org/}%
\providecommand \selectlanguage [0]{\@gobble}%
\providecommand \bibinfo  [0]{\@secondoftwo}%
\providecommand \bibfield  [0]{\@secondoftwo}%
\providecommand \translation [1]{[#1]}%
\providecommand \BibitemOpen [0]{}%
\providecommand \bibitemStop [0]{}%
\providecommand \bibitemNoStop [0]{.\EOS\space}%
\providecommand \EOS [0]{\spacefactor3000\relax}%
\providecommand \BibitemShut  [1]{\csname bibitem#1\endcsname}%
\let\auto@bib@innerbib\@empty
%</preamble>
\bibitem [{\citenamefont {Kochen}\ and\ \citenamefont
  {Specker}(1967)}]{kochen1}%
  \BibitemOpen
  \bibfield  {author} {\bibinfo {author} {\bibfnamefont {S.}~\bibnamefont
  {Kochen}}\ and\ \bibinfo {author} {\bibfnamefont {E.~P.}\ \bibnamefont
  {Specker}},\ }\bibfield  {title} {\bibinfo {title} {The problem of hidden
  variables in quantum mechanics},\ }\href
  {https://doi.org/10.1512/iumj.1968.17.17004} {\bibfield  {journal} {\bibinfo
  {journal} {Journal of Mathematics and Mechanics (now Indiana University
  Mathematics Journal)}\ }\textbf {\bibinfo {volume} {17}},\ \bibinfo {pages}
  {59} (\bibinfo {year} {1967})}\BibitemShut {NoStop}%
\bibitem [{\citenamefont {Boole}(1862)}]{Boole-62}%
  \BibitemOpen
  \bibfield  {author} {\bibinfo {author} {\bibfnamefont {G.}~\bibnamefont
  {Boole}},\ }\bibfield  {title} {\bibinfo {title} {On the theory of
  probabilities},\ }\href {https://doi.org/10.1098/rstl.1862.0015} {\bibfield
  {journal} {\bibinfo  {journal} {Philosophical Transactions of the Royal
  Society of London}\ }\textbf {\bibinfo {volume} {152}},\ \bibinfo {pages}
  {225} (\bibinfo {year} {1862})}\BibitemShut {NoStop}%
\bibitem [{\citenamefont {Schr{\"{o}}dinger}(1935)}]{schrodinger-gwsidqm2}%
  \BibitemOpen
  \bibfield  {author} {\bibinfo {author} {\bibfnamefont {E.}~\bibnamefont
  {Schr{\"{o}}dinger}},\ }\bibfield  {title} {\bibinfo {title} {{D}ie
  gegenw\"artige {S}ituation in der {Q}uantenmechanik},\ }\href
  {https://doi.org/10.1007/BF01491914} {\bibfield  {journal} {\bibinfo
  {journal} {Naturwissenschaften}\ }\textbf {\bibinfo {volume} {23}},\ \bibinfo
  {pages} {823} (\bibinfo {year} {1935})}\BibitemShut {NoStop}%
\bibitem [{\citenamefont {Bohr}(1949)}]{bohr-1949}%
  \BibitemOpen
  \bibfield  {author} {\bibinfo {author} {\bibfnamefont {N.}~\bibnamefont
  {Bohr}},\ }\bibfield  {title} {\bibinfo {title} {Discussion with {E}instein
  on epistemological problems in atomic physics},\ }in\ \href
  {https://doi.org/10.1016/S1876-0503(08)70379-7} {\emph {\bibinfo {booktitle}
  {{A}lbert {E}instein: Philosopher-Scientist}}},\ \bibinfo {editor} {edited
  by\ \bibinfo {editor} {\bibfnamefont {P.~A.}\ \bibnamefont {Schilpp}}}\
  (\bibinfo  {publisher} {The Library of Living Philosophers},\ \bibinfo
  {address} {Evanston, Ill.},\ \bibinfo {year} {1949})\ pp.\ \bibinfo {pages}
  {200--241}\BibitemShut {NoStop}%
\bibitem [{\citenamefont {Khrennikov}(2017)}]{Khrennikov2017}%
  \BibitemOpen
  \bibfield  {author} {\bibinfo {author} {\bibfnamefont {A.}~\bibnamefont
  {Khrennikov}},\ }\bibfield  {title} {\bibinfo {title} {{B}ohr against {B}ell:
  complementarity versus nonlocality},\ }\href
  {https://doi.org/10.1515/phys-2017-0086} {\bibfield  {journal} {\bibinfo
  {journal} {Open Physics}\ }\textbf {\bibinfo {volume} {15}},\ \bibinfo
  {pages} {734} (\bibinfo {year} {2017})}\BibitemShut {NoStop}%
\bibitem [{\citenamefont {Jaeger}(2019)}]{Jaeger2019}%
  \BibitemOpen
  \bibfield  {author} {\bibinfo {author} {\bibfnamefont {G.}~\bibnamefont
  {Jaeger}},\ }\bibfield  {title} {\bibinfo {title} {Quantum contextuality in
  the copenhagen approach},\ }\href {https://doi.org/10.1098/rsta.2019.0025}
  {\bibfield  {journal} {\bibinfo  {journal} {Philosophical Transactions of the
  Royal Society A: Mathematical, Physical and Engineering Sciences}\ }\textbf
  {\bibinfo {volume} {377}},\ \bibinfo {pages} {20190025} (\bibinfo {year}
  {2019})}\BibitemShut {NoStop}%
\bibitem [{\citenamefont {Khrennikov}(2009{\natexlab{a}})}]{Khrennikov2009a}%
  \BibitemOpen
  \bibfield  {author} {\bibinfo {author} {\bibfnamefont {A.}~\bibnamefont
  {Khrennikov}},\ }\href {https://doi.org/10.1515/9783110213195} {\emph
  {\bibinfo {title} {Interpretations of Probability}}},\ \bibinfo {edition}
  {2nd}\ ed.\ (\bibinfo  {publisher} {Walter de Gruyter},\ \bibinfo {address}
  {Berlin, New York},\ \bibinfo {year} {2009})\BibitemShut {NoStop}%
\bibitem [{\citenamefont {Khrennikov}(2009{\natexlab{b}})}]{Khrennikov2009b}%
  \BibitemOpen
  \bibfield  {author} {\bibinfo {author} {\bibfnamefont {A.}~\bibnamefont
  {Khrennikov}},\ }\href {https://doi.org/10.1007/978-1-4020-9593-1} {\emph
  {\bibinfo {title} {Contextual Approach to Quantum Formalism}}},\ \bibinfo
  {series} {Fundamental Theories of Physics}, Vol.\ \bibinfo {volume} {160}\
  (\bibinfo  {publisher} {Springer Science + Business Media B.V.},\ \bibinfo
  {year} {2009})\BibitemShut {NoStop}%
\bibitem [{\citenamefont {London}\ and\ \citenamefont
  {Bauer}(1939)}]{london-Bauer-1939}%
  \BibitemOpen
  \bibfield  {author} {\bibinfo {author} {\bibfnamefont {F.}~\bibnamefont
  {London}}\ and\ \bibinfo {author} {\bibfnamefont {E.}~\bibnamefont {Bauer}},\
  }\href@noop {} {\emph {\bibinfo {title} {La theorie de l'observation en
  m\'ecanique quantique; {N}o.~775 of Actualit\'es scientifiques et
  industrielles: Expos\'es de physique g\'en\'erale, publi\'es sous la
  direction de {P}aul {L}angevin}}}\ (\bibinfo  {publisher} {Hermann},\
  \bibinfo {address} {Paris},\ \bibinfo {year} {1939})\ \bibinfo {note}
  {english translation in~\cite{london-Bauer-1983}}\BibitemShut {NoStop}%
\bibitem [{\citenamefont {London}\ and\ \citenamefont
  {Bauer}(1983)}]{london-Bauer-1983}%
  \BibitemOpen
  \bibfield  {author} {\bibinfo {author} {\bibfnamefont {F.}~\bibnamefont
  {London}}\ and\ \bibinfo {author} {\bibfnamefont {E.}~\bibnamefont {Bauer}},\
  }\bibfield  {title} {\bibinfo {title} {The theory of observation in quantum
  mechanics},\ }in\ \href@noop {} {\emph {\bibinfo {booktitle} {Quantum Theory
  and Measurement}}},\ \bibinfo {editor} {edited by\ \bibinfo {editor}
  {\bibfnamefont {J.~A.}\ \bibnamefont {Wheeler}}\ and\ \bibinfo {editor}
  {\bibfnamefont {W.~H.}\ \bibnamefont {Zurek}}}\ (\bibinfo  {publisher}
  {Princeton University Press},\ \bibinfo {address} {Princeton, NJ},\ \bibinfo
  {year} {1983})\ pp.\ \bibinfo {pages} {217--259},\ \bibinfo {note}
  {consolidated translation of French
  original~\cite{london-Bauer-1939}}\BibitemShut {NoStop}%
\bibitem [{\citenamefont {Zeilinger}(1999)}]{zeil-99}%
  \BibitemOpen
  \bibfield  {author} {\bibinfo {author} {\bibfnamefont {A.}~\bibnamefont
  {Zeilinger}},\ }\bibfield  {title} {\bibinfo {title} {A foundational
  principle for quantum mechanics},\ }\href
  {https://doi.org/10.1023/A:1018820410908} {\bibfield  {journal} {\bibinfo
  {journal} {Foundations of Physics}\ }\textbf {\bibinfo {volume} {29}},\
  \bibinfo {pages} {631} (\bibinfo {year} {1999})}\BibitemShut {NoStop}%
\bibitem [{\citenamefont {Bell}(1966)}]{bell-66}%
  \BibitemOpen
  \bibfield  {author} {\bibinfo {author} {\bibfnamefont {J.~S.}\ \bibnamefont
  {Bell}},\ }\bibfield  {title} {\bibinfo {title} {On the problem of hidden
  variables in quantum mechanics},\ }\href
  {https://doi.org/10.1103/RevModPhys.38.447} {\bibfield  {journal} {\bibinfo
  {journal} {Reviews of Modern Physics}\ }\textbf {\bibinfo {volume} {38}},\
  \bibinfo {pages} {447} (\bibinfo {year} {1966})}\BibitemShut {NoStop}%
\bibitem [{\citenamefont {Hertz}(1894)}]{hertz-94}%
  \BibitemOpen
  \bibfield  {author} {\bibinfo {author} {\bibfnamefont {H.}~\bibnamefont
  {Hertz}},\ }\href {https://archive.org/details/dieprinzipiende00hertgoog}
  {\emph {\bibinfo {title} {{P}rinzipien der {M}echanik}}}\ (\bibinfo
  {publisher} {Johann Ambrosius Barth (Arthur Meiner)},\ \bibinfo {address}
  {Leipzig},\ \bibinfo {year} {1894})\ \bibinfo {note} {mit einem Vorwort von
  H. von Helmholtz}\BibitemShut {NoStop}%
\bibitem [{\citenamefont {Specker}(1960)}]{specker-60}%
  \BibitemOpen
  \bibfield  {author} {\bibinfo {author} {\bibfnamefont {E.}~\bibnamefont
  {Specker}},\ }\bibfield  {title} {\bibinfo {title} {{D}ie {L}ogik nicht
  gleichzeitig entscheidbarer {A}ussagen},\ }\href
  {https://doi.org/10.1111/j.1746-8361.1960.tb00422.x} {\bibfield  {journal}
  {\bibinfo  {journal} {Dialectica}\ }\textbf {\bibinfo {volume} {14}},\
  \bibinfo {pages} {239} (\bibinfo {year} {1960})},\ \bibinfo {note} {english
  translation at {https://arxiv.org/abs/1103.4537}},\ \Eprint
  {https://arxiv.org/abs/arXiv:1103.4537} {arXiv:1103.4537} \BibitemShut
  {NoStop}%
\bibitem [{\citenamefont {Abbott}\ \emph {et~al.}(2015)\citenamefont {Abbott},
  \citenamefont {Calude},\ and\ \citenamefont {Svozil}}]{2015-AnalyticKS}%
  \BibitemOpen
  \bibfield  {author} {\bibinfo {author} {\bibfnamefont {A.~A.}\ \bibnamefont
  {Abbott}}, \bibinfo {author} {\bibfnamefont {C.~S.}\ \bibnamefont {Calude}},\
  and\ \bibinfo {author} {\bibfnamefont {K.}~\bibnamefont {Svozil}},\
  }\bibfield  {title} {\bibinfo {title} {A variant of the {K}ochen-{S}pecker
  theorem localising value indefiniteness},\ }\href
  {https://doi.org/10.1063/1.4931658} {\bibfield  {journal} {\bibinfo
  {journal} {Journal of Mathematical Physics}\ }\textbf {\bibinfo {volume}
  {56}},\ \bibinfo {eid} {102201} (\bibinfo {year} {2015})},\ \Eprint
  {https://arxiv.org/abs/arXiv:1503.01985} {arXiv:1503.01985} \BibitemShut
  {NoStop}%
\bibitem [{\citenamefont {Budroni}\ \emph {et~al.}(2021)\citenamefont
  {Budroni}, \citenamefont {Cabello}, \citenamefont {G\"uhne}, \citenamefont
  {Kleinmann},\ and\ \citenamefont {Larsson}}]{cabello2021contextuality}%
  \BibitemOpen
  \bibfield  {author} {\bibinfo {author} {\bibfnamefont {C.}~\bibnamefont
  {Budroni}}, \bibinfo {author} {\bibfnamefont {A.}~\bibnamefont {Cabello}},
  \bibinfo {author} {\bibfnamefont {O.}~\bibnamefont {G\"uhne}}, \bibinfo
  {author} {\bibfnamefont {M.}~\bibnamefont {Kleinmann}},\ and\ \bibinfo
  {author} {\bibfnamefont {J.-A.}\ \bibnamefont {Larsson}},\ }\href
  {https://doi.org/10.48550/arXiv.2102.13036} {\bibinfo {title} {Quantum
  contextuality}} (\bibinfo {year} {2021}),\ \Eprint
  {https://arxiv.org/abs/2102.13036} {arXiv:2102.13036 [quant-ph]} \BibitemShut
  {NoStop}%
\bibitem [{\citenamefont {Cabello}(2008)}]{cabello:210401}%
  \BibitemOpen
  \bibfield  {author} {\bibinfo {author} {\bibfnamefont {A.}~\bibnamefont
  {Cabello}},\ }\bibfield  {title} {\bibinfo {title} {Experimentally testable
  state-independent quantum contextuality},\ }\href
  {https://doi.org/10.1103/PhysRevLett.101.210401} {\bibfield  {journal}
  {\bibinfo  {journal} {Physical Review Letters}\ }\textbf {\bibinfo {volume}
  {101}},\ \bibinfo {eid} {210401} (\bibinfo {year} {2008})},\ \Eprint
  {https://arxiv.org/abs/arXiv:0808.2456} {arXiv:0808.2456} \BibitemShut
  {NoStop}%
\bibitem [{\citenamefont {Cabello}\ \emph {et~al.}(2018)\citenamefont
  {Cabello}, \citenamefont {Portillo}, \citenamefont {Sol\'{i}s},\ and\
  \citenamefont {Svozil}}]{2018-minimalYIYS}%
  \BibitemOpen
  \bibfield  {author} {\bibinfo {author} {\bibfnamefont {A.}~\bibnamefont
  {Cabello}}, \bibinfo {author} {\bibfnamefont {J.~R.}\ \bibnamefont
  {Portillo}}, \bibinfo {author} {\bibfnamefont {A.}~\bibnamefont
  {Sol\'{i}s}},\ and\ \bibinfo {author} {\bibfnamefont {K.}~\bibnamefont
  {Svozil}},\ }\bibfield  {title} {\bibinfo {title} {Minimal true-implies-false
  and true-implies-true sets of propositions in noncontextual hidden-variable
  theories},\ }\href {https://doi.org/10.1103/PhysRevA.98.012106} {\bibfield
  {journal} {\bibinfo  {journal} {Physical Review A}\ }\textbf {\bibinfo
  {volume} {98}},\ \bibinfo {pages} {012106} (\bibinfo {year} {2018})},\
  \Eprint {https://arxiv.org/abs/arXiv:1805.00796} {arXiv:1805.00796}
  \BibitemShut {NoStop}%
\bibitem [{\citenamefont {Gleason}(1957)}]{Gleason}%
  \BibitemOpen
  \bibfield  {author} {\bibinfo {author} {\bibfnamefont {A.~M.}\ \bibnamefont
  {Gleason}},\ }\bibfield  {title} {\bibinfo {title} {Measures on the closed
  subspaces of a {H}ilbert space},\ }\href
  {https://doi.org/10.1512/iumj.1957.6.56050} {\bibfield  {journal} {\bibinfo
  {journal} {Journal of Mathematics and Mechanics (now Indiana University
  Mathematics Journal)}\ }\textbf {\bibinfo {volume} {6}},\ \bibinfo {pages}
  {885} (\bibinfo {year} {1957})}\BibitemShut {NoStop}%
\bibitem [{\citenamefont {Zierler}\ and\ \citenamefont
  {Schlessinger}(1965)}]{ZirlSchl-65}%
  \BibitemOpen
  \bibfield  {author} {\bibinfo {author} {\bibfnamefont {N.}~\bibnamefont
  {Zierler}}\ and\ \bibinfo {author} {\bibfnamefont {M.}~\bibnamefont
  {Schlessinger}},\ }\bibfield  {title} {\bibinfo {title} {Boolean embeddings
  of orthomodular sets and quantum logic},\ }\href
  {https://doi.org/10.1215/S0012-7094-65-03224-2} {\bibfield  {journal}
  {\bibinfo  {journal} {Duke Mathematical Journal}\ }\textbf {\bibinfo {volume}
  {32}},\ \bibinfo {pages} {251} (\bibinfo {year} {1965})},\ \bibinfo {note}
  {reprinted in Ref.~\cite{Zierler1975}}\BibitemShut {NoStop}%
\bibitem [{\citenamefont {Kamber}(1965)}]{kamber65}%
  \BibitemOpen
  \bibfield  {author} {\bibinfo {author} {\bibfnamefont {F.}~\bibnamefont
  {Kamber}},\ }\bibfield  {title} {\bibinfo {title} {Zweiwertige
  {W}ahrscheinlichkeitsfunktionen auf orthokomplement{\"{a}}ren
  {V}erb{\"{a}}nden},\ }\href {https://doi.org/10.1007/BF01359975} {\bibfield
  {journal} {\bibinfo  {journal} {Mathematische Annalen}\ }\textbf {\bibinfo
  {volume} {158}},\ \bibinfo {pages} {158} (\bibinfo {year}
  {1965})}\BibitemShut {NoStop}%
\bibitem [{\citenamefont {Halmos}(1958)}]{halmos-vs}%
  \BibitemOpen
  \bibfield  {author} {\bibinfo {author} {\bibfnamefont {P.~R.}\ \bibnamefont
  {Halmos}},\ }\href {https://doi.org/10.1007/978-1-4612-6387-6} {\emph
  {\bibinfo {title} {Finite-Dimensional Vector Spaces}}},\ Undergraduate Texts
  in Mathematics\ (\bibinfo  {publisher} {Springer},\ \bibinfo {address} {New
  York},\ \bibinfo {year} {1958})\BibitemShut {NoStop}%
\bibitem [{\citenamefont {Svozil}(2005)}]{svozil-2001-eua}%
  \BibitemOpen
  \bibfield  {author} {\bibinfo {author} {\bibfnamefont {K.}~\bibnamefont
  {Svozil}},\ }\bibfield  {title} {\bibinfo {title} {Logical equivalence
  between generalized urn models and finite automata},\ }\href
  {https://doi.org/10.1007/s10773-005-7052-0} {\bibfield  {journal} {\bibinfo
  {journal} {International Journal of Theoretical Physics}\ }\textbf {\bibinfo
  {volume} {44}},\ \bibinfo {pages} {745} (\bibinfo {year} {2005})},\ \Eprint
  {https://arxiv.org/abs/arXiv:quant-ph/0209136} {arXiv:quant-ph/0209136}
  \BibitemShut {NoStop}%
\bibitem [{\citenamefont {Wright}(1990)}]{wright}%
  \BibitemOpen
  \bibfield  {author} {\bibinfo {author} {\bibfnamefont {R.}~\bibnamefont
  {Wright}},\ }\bibfield  {title} {\bibinfo {title} {Generalized urn models},\
  }\href {https://doi.org/10.1007/BF01889696} {\bibfield  {journal} {\bibinfo
  {journal} {Foundations of Physics}\ }\textbf {\bibinfo {volume} {20}},\
  \bibinfo {pages} {881} (\bibinfo {year} {1990})}\BibitemShut {NoStop}%
\bibitem [{\citenamefont {Moore}(1956)}]{e-f-moore}%
  \BibitemOpen
  \bibfield  {author} {\bibinfo {author} {\bibfnamefont {E.~F.}\ \bibnamefont
  {Moore}},\ }\bibfield  {title} {\bibinfo {title} {Gedanken-experiments on
  sequential machines},\ }in\ \href {https://doi.org/10.1515/9781400882618-006}
  {\emph {\bibinfo {booktitle} {Automata Studies. {(AM-34)}}}},\ \bibinfo
  {editor} {edited by\ \bibinfo {editor} {\bibfnamefont {C.~E.}\ \bibnamefont
  {Shannon}}\ and\ \bibinfo {editor} {\bibfnamefont {J.}~\bibnamefont
  {McCarthy}}}\ (\bibinfo  {publisher} {Princeton University Press},\ \bibinfo
  {address} {Princeton, NJ},\ \bibinfo {year} {1956})\ pp.\ \bibinfo {pages}
  {129--153}\BibitemShut {NoStop}%
\bibitem [{\citenamefont {Schaller}\ and\ \citenamefont
  {Svozil}(1996)}]{schaller-96}%
  \BibitemOpen
  \bibfield  {author} {\bibinfo {author} {\bibfnamefont {M.}~\bibnamefont
  {Schaller}}\ and\ \bibinfo {author} {\bibfnamefont {K.}~\bibnamefont
  {Svozil}},\ }\bibfield  {title} {\bibinfo {title} {Automaton logic},\ }\href
  {https://doi.org/10.1007/BF02302381} {\bibfield  {journal} {\bibinfo
  {journal} {International Journal of Theoretical Physics}\ }\textbf {\bibinfo
  {volume} {35}},\ \bibinfo {pages} {911} (\bibinfo {year} {1996})}\BibitemShut
  {NoStop}%
\bibitem [{\citenamefont {Svozil}(2020{\natexlab{a}})}]{svozil-2018-b}%
  \BibitemOpen
  \bibfield  {author} {\bibinfo {author} {\bibfnamefont {K.}~\bibnamefont
  {Svozil}},\ }\bibfield  {title} {\bibinfo {title} {Faithful orthogonal
  representations of graphs from partition logics},\ }\href
  {https://doi.org/10.1007/s00500-019-04425-1} {\bibfield  {journal} {\bibinfo
  {journal} {Soft Computing}\ }\textbf {\bibinfo {volume} {24}},\ \bibinfo
  {pages} {10239} (\bibinfo {year} {2020}{\natexlab{a}})},\ \Eprint
  {https://arxiv.org/abs/arXiv:1810.10423} {arXiv:1810.10423} \BibitemShut
  {NoStop}%
\bibitem [{\citenamefont {Froissart}(1981)}]{froissart-81}%
  \BibitemOpen
  \bibfield  {author} {\bibinfo {author} {\bibfnamefont {M.}~\bibnamefont
  {Froissart}},\ }\bibfield  {title} {\bibinfo {title} {Constructive
  generalization of {B}ell's inequalities},\ }\href
  {https://doi.org/10.1007/BF02903286} {\bibfield  {journal} {\bibinfo
  {journal} {Il Nuovo Cimento B (11, 1971-1996)}\ }\textbf {\bibinfo {volume}
  {64}},\ \bibinfo {pages} {241} (\bibinfo {year} {1981})}\BibitemShut
  {NoStop}%
\bibitem [{\citenamefont {Pitowsky}(1986)}]{pitowsky-86}%
  \BibitemOpen
  \bibfield  {author} {\bibinfo {author} {\bibfnamefont {I.}~\bibnamefont
  {Pitowsky}},\ }\bibfield  {title} {\bibinfo {title} {The range of quantum
  probability},\ }\href {https://doi.org/10.1063/1.527066} {\bibfield
  {journal} {\bibinfo  {journal} {Journal of Mathematical Physics}\ }\textbf
  {\bibinfo {volume} {27}},\ \bibinfo {pages} {1556} (\bibinfo {year}
  {1986})}\BibitemShut {NoStop}%
\bibitem [{\citenamefont {Foulis}\ and\ \citenamefont
  {Randall}(1976)}]{Foulis1976}%
  \BibitemOpen
  \bibfield  {author} {\bibinfo {author} {\bibfnamefont {D.~J.}\ \bibnamefont
  {Foulis}}\ and\ \bibinfo {author} {\bibfnamefont {C.~H.}\ \bibnamefont
  {Randall}},\ }\bibfield  {title} {\bibinfo {title} {Empirical logic and
  quantum mechanics},\ }in\ \href
  {https://doi.org/10.1007/978-94-010-9466-5\_5} {\emph {\bibinfo {booktitle}
  {Logic and Probability in Quantum Mechanics}}},\ \bibinfo {editor} {edited
  by\ \bibinfo {editor} {\bibfnamefont {P.}~\bibnamefont {Suppes}}}\ (\bibinfo
  {publisher} {Springer Netherlands},\ \bibinfo {address} {Dordrecht},\
  \bibinfo {year} {1976})\ pp.\ \bibinfo {pages} {73--103}\BibitemShut
  {NoStop}%
\bibitem [{\citenamefont {Birkhoff}\ and\ \citenamefont {{von
  Neumann}}(1936)}]{birkhoff-36}%
  \BibitemOpen
  \bibfield  {author} {\bibinfo {author} {\bibfnamefont {G.}~\bibnamefont
  {Birkhoff}}\ and\ \bibinfo {author} {\bibfnamefont {J.}~\bibnamefont {{von
  Neumann}}},\ }\bibfield  {title} {\bibinfo {title} {The logic of quantum
  mechanics},\ }\href {https://doi.org/10.2307/1968621} {\bibfield  {journal}
  {\bibinfo  {journal} {Annals of Mathematics}\ }\textbf {\bibinfo {volume}
  {37}},\ \bibinfo {pages} {823} (\bibinfo {year} {1936})}\BibitemShut
  {NoStop}%
\bibitem [{\citenamefont {Suppes}\ and\ \citenamefont
  {Zanotti}(1981)}]{Suppes-81}%
  \BibitemOpen
  \bibfield  {author} {\bibinfo {author} {\bibfnamefont {P.}~\bibnamefont
  {Suppes}}\ and\ \bibinfo {author} {\bibfnamefont {M.}~\bibnamefont
  {Zanotti}},\ }\bibfield  {title} {\bibinfo {title} {When are probabilistic
  explanations possible?},\ }\href {https://doi.org/10.1007/BF01063886}
  {\bibfield  {journal} {\bibinfo  {journal} {Synthese}\ }\textbf {\bibinfo
  {volume} {48}},\ \bibinfo {pages} {191} (\bibinfo {year} {1981})}\BibitemShut
  {NoStop}%
\bibitem [{\citenamefont {Avis}\ \emph {et~al.}(2005)\citenamefont {Avis},
  \citenamefont {Imai}, \citenamefont {Ito},\ and\ \citenamefont
  {Sasaki}}]{Avis2005}%
  \BibitemOpen
  \bibfield  {author} {\bibinfo {author} {\bibfnamefont {D.}~\bibnamefont
  {Avis}}, \bibinfo {author} {\bibfnamefont {H.}~\bibnamefont {Imai}}, \bibinfo
  {author} {\bibfnamefont {T.}~\bibnamefont {Ito}},\ and\ \bibinfo {author}
  {\bibfnamefont {Y.}~\bibnamefont {Sasaki}},\ }\bibfield  {title} {\bibinfo
  {title} {Two-party {B}ell inequalities derived from combinatorics via
  triangular elimination},\ }\href
  {https://doi.org/10.1088/0305-4470/38/50/007} {\bibfield  {journal} {\bibinfo
   {journal} {Journal of Physics A: Mathematical and General}\ }\textbf
  {\bibinfo {volume} {38}},\ \bibinfo {pages} {10971} (\bibinfo {year}
  {2005})}\BibitemShut {NoStop}%
\bibitem [{\citenamefont {Kochen}\ and\ \citenamefont
  {Specker}(1965)}]{kochen2}%
  \BibitemOpen
  \bibfield  {author} {\bibinfo {author} {\bibfnamefont {S.}~\bibnamefont
  {Kochen}}\ and\ \bibinfo {author} {\bibfnamefont {E.~P.}\ \bibnamefont
  {Specker}},\ }\bibfield  {title} {\bibinfo {title} {Logical structures
  arising in quantum theory},\ }in\ \href
  {https://doi.org/978-3-0348-9259-9_19} {\emph {\bibinfo {booktitle} {The
  Theory of Models, {P}roceedings of the 1963 International Symposium at
  {B}erkeley}}},\ \bibinfo {editor} {edited by\ \bibinfo {editor}
  {\bibfnamefont {J.~W.}\ \bibnamefont {Addison}}, \bibinfo {editor}
  {\bibfnamefont {L.}~\bibnamefont {Henkin}},\ and\ \bibinfo {editor}
  {\bibfnamefont {A.}~\bibnamefont {Tarski}}}\ (\bibinfo  {publisher} {North
  Holland},\ \bibinfo {address} {Amsterdam, New York, Oxford},\ \bibinfo {year}
  {1965})\ pp.\ \bibinfo {pages} {177--189},\ \bibinfo {note} {reprinted in
  Ref.~\cite[pp.~209-221]{specker-ges}}\BibitemShut {NoStop}%
\bibitem [{\citenamefont {Svozil}(2001)}]{svozil-2001-cesena}%
  \BibitemOpen
  \bibfield  {author} {\bibinfo {author} {\bibfnamefont {K.}~\bibnamefont
  {Svozil}},\ }\href {https://arxiv.org/abs/quant-ph/0012066} {\bibinfo {title}
  {On generalized probabilities: correlation polytopes for automaton logic and
  generalized urn models, extensions of quantum mechanics and parameter
  cheats}} (\bibinfo {year} {2001}),\ \Eprint
  {https://arxiv.org/abs/arXiv:quant-ph/0012066} {arXiv:quant-ph/0012066}
  \BibitemShut {NoStop}%
\bibitem [{\citenamefont {Klyachko}\ \emph {et~al.}(2008)\citenamefont
  {Klyachko}, \citenamefont {Can}, \citenamefont
  {Binicio\ifmmode~\breve{g}\else \u{g}\fi{}lu},\ and\ \citenamefont
  {Shumovsky}}]{Klyachko-2008}%
  \BibitemOpen
  \bibfield  {author} {\bibinfo {author} {\bibfnamefont {A.~A.}\ \bibnamefont
  {Klyachko}}, \bibinfo {author} {\bibfnamefont {M.~A.}\ \bibnamefont {Can}},
  \bibinfo {author} {\bibfnamefont {S.}~\bibnamefont
  {Binicio\ifmmode~\breve{g}\else \u{g}\fi{}lu}},\ and\ \bibinfo {author}
  {\bibfnamefont {A.~S.}\ \bibnamefont {Shumovsky}},\ }\bibfield  {title}
  {\bibinfo {title} {Simple test for hidden variables in spin-1 systems},\
  }\href {https://doi.org/10.1103/PhysRevLett.101.020403} {\bibfield  {journal}
  {\bibinfo  {journal} {Physical Review Letters}\ }\textbf {\bibinfo {volume}
  {101}},\ \bibinfo {pages} {020403} (\bibinfo {year} {2008})},\ \Eprint
  {https://arxiv.org/abs/arXiv:0706.0126} {arXiv:0706.0126} \BibitemShut
  {NoStop}%
\bibitem [{\citenamefont {Bub}\ and\ \citenamefont {Stairs}(2009)}]{Bub-2009}%
  \BibitemOpen
  \bibfield  {author} {\bibinfo {author} {\bibfnamefont {J.}~\bibnamefont
  {Bub}}\ and\ \bibinfo {author} {\bibfnamefont {A.}~\bibnamefont {Stairs}},\
  }\bibfield  {title} {\bibinfo {title} {Contextuality and nonlocality in `no
  signaling' theories},\ }\href {https://doi.org/10.1007/s10701-009-9307-8}
  {\bibfield  {journal} {\bibinfo  {journal} {Foundations of Physics}\ }\textbf
  {\bibinfo {volume} {39}},\ \bibinfo {pages} {690} (\bibinfo {year} {2009})},\
  \Eprint {https://arxiv.org/abs/arXiv:0903.1462} {arXiv:0903.1462}
  \BibitemShut {NoStop}%
\bibitem [{\citenamefont {Svozil}(2021{\natexlab{a}})}]{svozil-2020-ex}%
  \BibitemOpen
  \bibfield  {author} {\bibinfo {author} {\bibfnamefont {K.}~\bibnamefont
  {Svozil}},\ }\bibfield  {title} {\bibinfo {title} {Quantum violation of the
  {S}uppes-{Z}anotti inequalities and ``contextuality''},\ }\href
  {https://doi.org/10.1007/s10773-021-04850-9} {\bibfield  {journal} {\bibinfo
  {journal} {International Journal of Theoretical Physics}\ }\textbf {\bibinfo
  {volume} {60}},\ \bibinfo {pages} {2300} (\bibinfo {year}
  {2021}{\natexlab{a}})},\ \Eprint {https://arxiv.org/abs/arXiv:2101.10167}
  {arXiv:2101.10167} \BibitemShut {NoStop}%
\bibitem [{\citenamefont {Clauser}\ \emph {et~al.}(1969)\citenamefont
  {Clauser}, \citenamefont {Horne}, \citenamefont {Shimony},\ and\
  \citenamefont {Holt}}]{chsh}%
  \BibitemOpen
  \bibfield  {author} {\bibinfo {author} {\bibfnamefont {J.~F.}\ \bibnamefont
  {Clauser}}, \bibinfo {author} {\bibfnamefont {M.~A.}\ \bibnamefont {Horne}},
  \bibinfo {author} {\bibfnamefont {A.}~\bibnamefont {Shimony}},\ and\ \bibinfo
  {author} {\bibfnamefont {R.~A.}\ \bibnamefont {Holt}},\ }\bibfield  {title}
  {\bibinfo {title} {Proposed experiment to test local hidden-variable
  theories},\ }\href {https://doi.org/10.1103/PhysRevLett.23.880} {\bibfield
  {journal} {\bibinfo  {journal} {Physical Review Letters}\ }\textbf {\bibinfo
  {volume} {23}},\ \bibinfo {pages} {880} (\bibinfo {year} {1969})}\BibitemShut
  {NoStop}%
\bibitem [{\citenamefont {Peres}(1978)}]{peres222}%
  \BibitemOpen
  \bibfield  {author} {\bibinfo {author} {\bibfnamefont {A.}~\bibnamefont
  {Peres}},\ }\bibfield  {title} {\bibinfo {title} {Unperformed experiments
  have no results},\ }\href {https://doi.org/10.1119/1.11393} {\bibfield
  {journal} {\bibinfo  {journal} {American Journal of Physics}\ }\textbf
  {\bibinfo {volume} {46}},\ \bibinfo {pages} {745} (\bibinfo {year}
  {1978})}\BibitemShut {NoStop}%
\bibitem [{\citenamefont {Svozil}(2011)}]{svozil_2010-pc09}%
  \BibitemOpen
  \bibfield  {author} {\bibinfo {author} {\bibfnamefont {K.}~\bibnamefont
  {Svozil}},\ }\bibfield  {title} {\bibinfo {title} {Quantum value
  indefiniteness},\ }\href {https://doi.org/10.1007/s11047-010-9241-x}
  {\bibfield  {journal} {\bibinfo  {journal} {Natural Computing}\ }\textbf
  {\bibinfo {volume} {10}},\ \bibinfo {pages} {1371} (\bibinfo {year}
  {2011})},\ \Eprint {https://arxiv.org/abs/arXiv:1001.1436} {arXiv:1001.1436}
  \BibitemShut {NoStop}%
\bibitem [{\citenamefont {Svozil}(2012)}]{svozil-2011-enough}%
  \BibitemOpen
  \bibfield  {author} {\bibinfo {author} {\bibfnamefont {K.}~\bibnamefont
  {Svozil}},\ }\bibfield  {title} {\bibinfo {title} {How much contextuality?},\
  }\href {https://doi.org/10.1007/s11047-012-9318-9} {\bibfield  {journal}
  {\bibinfo  {journal} {Natural Computing}\ }\textbf {\bibinfo {volume} {11}},\
  \bibinfo {pages} {261} (\bibinfo {year} {2012})},\ \Eprint
  {https://arxiv.org/abs/arXiv:1103.3980} {arXiv:1103.3980} \BibitemShut
  {NoStop}%
\bibitem [{\citenamefont {Dzhafarov}\ \emph {et~al.}(2015)\citenamefont
  {Dzhafarov}, \citenamefont {Kujala},\ and\ \citenamefont
  {Larsson}}]{Dzhafarov-2015}%
  \BibitemOpen
  \bibfield  {author} {\bibinfo {author} {\bibfnamefont {E.~N.}\ \bibnamefont
  {Dzhafarov}}, \bibinfo {author} {\bibfnamefont {J.~V.}\ \bibnamefont
  {Kujala}},\ and\ \bibinfo {author} {\bibfnamefont {J.-A.~k.}\ \bibnamefont
  {Larsson}},\ }\bibfield  {title} {\bibinfo {title} {Contextuality in three
  types of quantum-mechanical systems},\ }\href
  {https://doi.org/10.1007/s10701-015-9882-9} {\bibfield  {journal} {\bibinfo
  {journal} {Foundations of Physics}\ }\textbf {\bibinfo {volume} {45}},\
  \bibinfo {pages} {762} (\bibinfo {year} {2015})}\BibitemShut {NoStop}%
\bibitem [{\citenamefont {Dzhafarov}\ \emph {et~al.}(2017)\citenamefont
  {Dzhafarov}, \citenamefont {Cervantes},\ and\ \citenamefont
  {Kujala}}]{Dzhafarov-2017}%
  \BibitemOpen
  \bibfield  {author} {\bibinfo {author} {\bibfnamefont {E.~N.}\ \bibnamefont
  {Dzhafarov}}, \bibinfo {author} {\bibfnamefont {V.~H.}\ \bibnamefont
  {Cervantes}},\ and\ \bibinfo {author} {\bibfnamefont {J.~V.}\ \bibnamefont
  {Kujala}},\ }\bibfield  {title} {\bibinfo {title} {Contextuality in canonical
  systems of random variables},\ }\href
  {https://doi.org/10.1098/rsta.2016.0389} {\bibfield  {journal} {\bibinfo
  {journal} {Philosophical Transactions of the Royal Society A: Mathematical,
  Physical and Engineering Sciences}\ }\textbf {\bibinfo {volume} {375}},\
  \bibinfo {pages} {20160389} (\bibinfo {year} {2017})},\ \Eprint
  {https://arxiv.org/abs/arXiv:1703.01252} {arXiv:1703.01252} \BibitemShut
  {NoStop}%
\bibitem [{\citenamefont {Kujala}\ and\ \citenamefont
  {Dzhafarov}(2019)}]{KujalaDzhafarov-2019}%
  \BibitemOpen
  \bibfield  {author} {\bibinfo {author} {\bibfnamefont {J.~V.}\ \bibnamefont
  {Kujala}}\ and\ \bibinfo {author} {\bibfnamefont {E.~N.}\ \bibnamefont
  {Dzhafarov}},\ }\bibfield  {title} {\bibinfo {title} {Measures of
  contextuality and non-contextuality},\ }\href
  {https://doi.org/10.1098/rsta.2019.0149} {\bibfield  {journal} {\bibinfo
  {journal} {Philosophical Transactions of the Royal Society A. Mathematical,
  Physical and Engineering Sciences}\ }\textbf {\bibinfo {volume} {377}},\
  \bibinfo {pages} {20190149, 16} (\bibinfo {year} {2019})}\BibitemShut
  {NoStop}%
\bibitem [{\citenamefont {Dzhafarov}\ \emph {et~al.}(2020)\citenamefont
  {Dzhafarov}, \citenamefont {Kujala},\ and\ \citenamefont
  {Cervantes}}]{Dzhafarov-19PhysRevA}%
  \BibitemOpen
  \bibfield  {author} {\bibinfo {author} {\bibfnamefont {E.~N.}\ \bibnamefont
  {Dzhafarov}}, \bibinfo {author} {\bibfnamefont {J.~V.}\ \bibnamefont
  {Kujala}},\ and\ \bibinfo {author} {\bibfnamefont {V.~H.}\ \bibnamefont
  {Cervantes}},\ }\bibfield  {title} {\bibinfo {title} {Contextuality and
  noncontextuality measures and generalized bell inequalities for cyclic
  systems},\ }\href {https://doi.org/10.1103/PhysRevA.101.042119} {\bibfield
  {journal} {\bibinfo  {journal} {Physical Review A}\ }\textbf {\bibinfo
  {volume} {101}},\ \bibinfo {pages} {042119} (\bibinfo {year}
  {2020})}\BibitemShut {NoStop}%
\bibitem [{\citenamefont {Einstein}\ \emph {et~al.}(1935)\citenamefont
  {Einstein}, \citenamefont {Podolsky},\ and\ \citenamefont {Rosen}}]{epr}%
  \BibitemOpen
  \bibfield  {author} {\bibinfo {author} {\bibfnamefont {A.}~\bibnamefont
  {Einstein}}, \bibinfo {author} {\bibfnamefont {B.}~\bibnamefont {Podolsky}},\
  and\ \bibinfo {author} {\bibfnamefont {N.}~\bibnamefont {Rosen}},\ }\bibfield
   {title} {\bibinfo {title} {Can quantum-mechanical description of physical
  reality be considered complete?},\ }\href
  {https://doi.org/10.1103/PhysRev.47.777} {\bibfield  {journal} {\bibinfo
  {journal} {Physical Review}\ }\textbf {\bibinfo {volume} {47}},\ \bibinfo
  {pages} {777} (\bibinfo {year} {1935})}\BibitemShut {NoStop}%
\bibitem [{\citenamefont {Garg}\ and\ \citenamefont {Mermin}(1984)}]{Garg1984}%
  \BibitemOpen
  \bibfield  {author} {\bibinfo {author} {\bibfnamefont {A.}~\bibnamefont
  {Garg}}\ and\ \bibinfo {author} {\bibfnamefont {D.~N.}\ \bibnamefont
  {Mermin}},\ }\bibfield  {title} {\bibinfo {title} {{F}arkas's lemma and the
  nature of reality: Statistical implications of quantum correlations},\ }\href
  {https://doi.org/10.1007/BF00741645} {\bibfield  {journal} {\bibinfo
  {journal} {Foundations of Physics}\ }\textbf {\bibinfo {volume} {14}},\
  \bibinfo {pages} {1} (\bibinfo {year} {1984})}\BibitemShut {NoStop}%
\bibitem [{\citenamefont {Ziegler}(1994)}]{ziegler}%
  \BibitemOpen
  \bibfield  {author} {\bibinfo {author} {\bibfnamefont {G.~M.}\ \bibnamefont
  {Ziegler}},\ }\href {https://doi.org/10.1007/978-1-4613-8431-1} {\emph
  {\bibinfo {title} {Lectures on Polytopes}}},\ \bibinfo {series} {Graduate
  Texts in Mathematics}, Vol.\ \bibinfo {volume} {152}\ (\bibinfo  {publisher}
  {Springer},\ \bibinfo {address} {New York},\ \bibinfo {year}
  {1994})\BibitemShut {NoStop}%
\bibitem [{\citenamefont {Schrijver}(1998)}]{Schrijver}%
  \BibitemOpen
  \bibfield  {author} {\bibinfo {author} {\bibfnamefont {A.}~\bibnamefont
  {Schrijver}},\ }\href
  {http://eu.wiley.com/WileyCDA/WileyTitle/productCd-0471982326.html} {\emph
  {\bibinfo {title} {Theory of Linear and Integer Programming}}},\ Wiley Series
  in Discrete Mathematics \& Optimization\ (\bibinfo  {publisher} {John Wiley
  \& Sons},\ \bibinfo {address} {New York, Toronto, London},\ \bibinfo {year}
  {1986, 1998})\BibitemShut {NoStop}%
\bibitem [{\citenamefont {Fukuda}(2014)}]{Fukuda-techrep}%
  \BibitemOpen
  \bibfield  {author} {\bibinfo {author} {\bibfnamefont {K.}~\bibnamefont
  {Fukuda}},\ }\bibfield  {title} {\bibinfo {title} {Frequently asked questions
  in polyhedral computation}} (\bibinfo {year} {2014}),\ \bibinfo {note}
  {accessed on July 29th, 2017}\BibitemShut {NoStop}%
\bibitem [{\citenamefont {Brody}(1989)}]{Brody-1989}%
  \BibitemOpen
  \bibfield  {author} {\bibinfo {author} {\bibfnamefont {T.~A.}\ \bibnamefont
  {Brody}},\ }\bibfield  {title} {\bibinfo {title} {The {S}uppes-{Z}anotti
  theorem and the {B}ell inequalities},\ }\href
  {https://rmf.smf.mx/ojs/rmf/article/view/2042/2010} {\bibfield  {journal}
  {\bibinfo  {journal} {Revista Mexicana de F\'{\i}sica}\ }\textbf {\bibinfo
  {volume} {35}},\ \bibinfo {pages} {170} (\bibinfo {year} {1989})}\BibitemShut
  {NoStop}%
\bibitem [{\citenamefont {Khrennikov}(2020)}]{Khrennikov2020}%
  \BibitemOpen
  \bibfield  {author} {\bibinfo {author} {\bibfnamefont {A.}~\bibnamefont
  {Khrennikov}},\ }\bibfield  {title} {\bibinfo {title} {Can there be given any
  meaning to contextuality without incompatibility?},\ }\bibfield  {journal}
  {\bibinfo  {journal} {International Journal of Theoretical Physics}\ }\href
  {https://doi.org/10.1007/s10773-020-04666-z} {10.1007/s10773-020-04666-z}
  (\bibinfo {year} {2020})\BibitemShut {NoStop}%
\bibitem [{\citenamefont {Filipp}\ and\ \citenamefont
  {Svozil}(2004)}]{filipp-svo-04-qpoly-prl}%
  \BibitemOpen
  \bibfield  {author} {\bibinfo {author} {\bibfnamefont {S.}~\bibnamefont
  {Filipp}}\ and\ \bibinfo {author} {\bibfnamefont {K.}~\bibnamefont
  {Svozil}},\ }\bibfield  {title} {\bibinfo {title} {Generalizing {T}sirelson's
  bound on {B}ell inequalities using a min-max principle},\ }\href
  {https://doi.org/10.1103/PhysRevLett.93.130407} {\bibfield  {journal}
  {\bibinfo  {journal} {Physical Review Letters}\ }\textbf {\bibinfo {volume}
  {93}},\ \bibinfo {pages} {130407} (\bibinfo {year} {2004})},\ \Eprint
  {https://arxiv.org/abs/arXiv:quant-ph/0403175} {arXiv:quant-ph/0403175}
  \BibitemShut {NoStop}%
\bibitem [{\citenamefont {Svozil}(2021{\natexlab{b}})}]{svozil-2020-hardy}%
  \BibitemOpen
  \bibfield  {author} {\bibinfo {author} {\bibfnamefont {K.}~\bibnamefont
  {Svozil}},\ }\bibfield  {title} {\bibinfo {title} {Extensions of {H}ardy-type
  true-implies-false gadgets to classically obtain indistinguishability},\
  }\href {https://doi.org/10.1103/PhysRevA.103.022204} {\bibfield  {journal}
  {\bibinfo  {journal} {Physical Review A}\ }\textbf {\bibinfo {volume}
  {103}},\ \bibinfo {pages} {022204} (\bibinfo {year} {2021}{\natexlab{b}})},\
  \Eprint {https://arxiv.org/abs/arXiv:2006.11396} {arXiv:2006.11396}
  \BibitemShut {NoStop}%
\bibitem [{\citenamefont {Ramanathan}\ \emph {et~al.}(2020)\citenamefont
  {Ramanathan}, \citenamefont {Rosicka}, \citenamefont {Horodecki},
  \citenamefont {Pironio}, \citenamefont {Horodecki},\ and\ \citenamefont
  {Horodecki}}]{Ramanathan-18}%
  \BibitemOpen
  \bibfield  {author} {\bibinfo {author} {\bibfnamefont {R.}~\bibnamefont
  {Ramanathan}}, \bibinfo {author} {\bibfnamefont {M.}~\bibnamefont {Rosicka}},
  \bibinfo {author} {\bibfnamefont {K.}~\bibnamefont {Horodecki}}, \bibinfo
  {author} {\bibfnamefont {S.}~\bibnamefont {Pironio}}, \bibinfo {author}
  {\bibfnamefont {M.}~\bibnamefont {Horodecki}},\ and\ \bibinfo {author}
  {\bibfnamefont {P.}~\bibnamefont {Horodecki}},\ }\href
  {https://doi.org/10.22331/q-2020-08-14-308} {\bibinfo {title} {Gadget
  structures in proofs of the {K}ochen-{S}pecker theorem}} (\bibinfo {year}
  {2020}),\ \Eprint {https://arxiv.org/abs/arXiv:1807.00113} {arXiv:1807.00113}
  \BibitemShut {NoStop}%
\bibitem [{\citenamefont {Svozil}(2018)}]{svozil-2018-whycontexts}%
  \BibitemOpen
  \bibfield  {author} {\bibinfo {author} {\bibfnamefont {K.}~\bibnamefont
  {Svozil}},\ }\bibfield  {title} {\bibinfo {title} {New forms of quantum value
  indefiniteness suggest that incompatible views on contexts are epistemic},\
  }\href {https://doi.org/10.3390/e20060406} {\bibfield  {journal} {\bibinfo
  {journal} {Entropy}\ }\textbf {\bibinfo {volume} {20}},\ \bibinfo {pages}
  {406(22)} (\bibinfo {year} {2018})},\ \Eprint
  {https://arxiv.org/abs/arXiv:1804.10030} {arXiv:1804.10030} \BibitemShut
  {NoStop}%
\bibitem [{\citenamefont {Lov\'asz}(1979)}]{lovasz-79}%
  \BibitemOpen
  \bibfield  {author} {\bibinfo {author} {\bibfnamefont {L.}~\bibnamefont
  {Lov\'asz}},\ }\bibfield  {title} {\bibinfo {title} {On the {S}hannon
  capacity of a graph},\ }\href {https://doi.org/10.1109/TIT.1979.1055985}
  {\bibfield  {journal} {\bibinfo  {journal} {IEEE Transactions on Information
  Theory}\ }\textbf {\bibinfo {volume} {25}},\ \bibinfo {pages} {1} (\bibinfo
  {year} {1979})}\BibitemShut {NoStop}%
\bibitem [{\citenamefont {Belinfante}(1973)}]{Belinfante-73}%
  \BibitemOpen
  \bibfield  {author} {\bibinfo {author} {\bibfnamefont {F.~J.}\ \bibnamefont
  {Belinfante}},\ }\bibfield  {title} {\bibinfo {title} {A survey of
  hidden-variables theories}\ }(\bibinfo  {publisher} {Pergamon Press,
  Elsevier},\ \bibinfo {address} {Oxford, New York},\ \bibinfo {year}
  {1973})\BibitemShut {NoStop}%
\bibitem [{\citenamefont {Redhead}(1987)}]{redhead}%
  \BibitemOpen
  \bibfield  {author} {\bibinfo {author} {\bibfnamefont {M.}~\bibnamefont
  {Redhead}},\ }\href
  {https://global.oup.com/academic/product/incompleteness-nonlocality-and-realism-9780198242383}
  {\emph {\bibinfo {title} {Incompleteness, Nonlocality, and Realism: A
  Prolegomenon to the Philosophy of Quantum Mechanics}}}\ (\bibinfo
  {publisher} {Clarendon Press},\ \bibinfo {address} {Oxford},\ \bibinfo {year}
  {1987})\BibitemShut {NoStop}%
\bibitem [{\citenamefont {Cabello}(1994)}]{cabello-1994}%
  \BibitemOpen
  \bibfield  {author} {\bibinfo {author} {\bibfnamefont {A.}~\bibnamefont
  {Cabello}},\ }\bibfield  {title} {\bibinfo {title} {A simple proof of the
  {K}ochen-{S}pecker theorem},\ }\href
  {https://doi.org/10.1088/0143-0807/15/4/004} {\bibfield  {journal} {\bibinfo
  {journal} {European Journal of Physics}\ }\textbf {\bibinfo {volume} {15}},\
  \bibinfo {pages} {179} (\bibinfo {year} {1994})}\BibitemShut {NoStop}%
\bibitem [{\citenamefont {Cabello}(2021)}]{cabello2020converting}%
  \BibitemOpen
  \bibfield  {author} {\bibinfo {author} {\bibfnamefont {A.}~\bibnamefont
  {Cabello}},\ }\bibfield  {title} {\bibinfo {title} {Converting contextuality
  into nonlocality},\ }\href {https://doi.org/10.1103/PhysRevLett.127.070401}
  {\bibfield  {journal} {\bibinfo  {journal} {Physical Review Letters}\
  }\textbf {\bibinfo {volume} {127}},\ \bibinfo {pages} {070401} (\bibinfo
  {year} {2021})},\ \Eprint {https://arxiv.org/abs/arXiv:2011.13790}
  {arXiv:2011.13790} \BibitemShut {NoStop}%
\bibitem [{\citenamefont {Tkadlec}(1998)}]{tkadlec-96}%
  \BibitemOpen
  \bibfield  {author} {\bibinfo {author} {\bibfnamefont {J.}~\bibnamefont
  {Tkadlec}},\ }\bibfield  {title} {\bibinfo {title} {{G}reechie diagrams of
  small quantum logics with small state spaces},\ }\href
  {https://doi.org/10.1023/A:1026646229896} {\bibfield  {journal} {\bibinfo
  {journal} {International Journal of Theoretical Physics}\ }\textbf {\bibinfo
  {volume} {37}},\ \bibinfo {pages} {203} (\bibinfo {year} {1998})}\BibitemShut
  {NoStop}%
\bibitem [{\citenamefont {Svozil}(2020{\natexlab{b}})}]{Svozil-2018-p}%
  \BibitemOpen
  \bibfield  {author} {\bibinfo {author} {\bibfnamefont {K.}~\bibnamefont
  {Svozil}},\ }\bibfield  {title} {\bibinfo {title} {Roots and (re)sources of
  value (in)definiteness versus contextuality},\ }in\ \href
  {https://doi.org/10.1007/978-3-030-34316-3\_24} {\emph {\bibinfo {booktitle}
  {Quantum, Probability, Logic: The Work and Influence of {I}tamar
  {P}itowsky}}},\ \bibinfo {series} {Jerusalem Studies in Philosophy and
  History of Science (JSPS)}, Vol.~\bibinfo {volume} {1},\ \bibinfo {editor}
  {edited by\ \bibinfo {editor} {\bibfnamefont {M.}~\bibnamefont {Hemmo}}\ and\
  \bibinfo {editor} {\bibfnamefont {O.}~\bibnamefont {Shenker}}}\ (\bibinfo
  {publisher} {Springer International Publishing},\ \bibinfo {address} {Cham},\
  \bibinfo {year} {2020})\ pp.\ \bibinfo {pages} {521--544},\ \Eprint
  {https://arxiv.org/abs/arXiv:1812.08646} {arXiv:1812.08646} \BibitemShut
  {NoStop}%
\bibitem [{\citenamefont {Pitowsky}(1998)}]{pitowsky:218}%
  \BibitemOpen
  \bibfield  {author} {\bibinfo {author} {\bibfnamefont {I.}~\bibnamefont
  {Pitowsky}},\ }\bibfield  {title} {\bibinfo {title} {Infinite and finite
  {G}leason's theorems and the logic of indeterminacy},\ }\href
  {https://doi.org/10.1063/1.532334} {\bibfield  {journal} {\bibinfo  {journal}
  {Journal of Mathematical Physics}\ }\textbf {\bibinfo {volume} {39}},\
  \bibinfo {pages} {218} (\bibinfo {year} {1998})}\BibitemShut {NoStop}%
\bibitem [{\citenamefont {Hrushovski}\ and\ \citenamefont
  {Pitowsky}(2004)}]{hru-pit-2003}%
  \BibitemOpen
  \bibfield  {author} {\bibinfo {author} {\bibfnamefont {E.}~\bibnamefont
  {Hrushovski}}\ and\ \bibinfo {author} {\bibfnamefont {I.}~\bibnamefont
  {Pitowsky}},\ }\bibfield  {title} {\bibinfo {title} {Generalizations of
  {K}ochen and {S}pecker's theorem and the effectiveness of {G}leason's
  theorem},\ }\href {https://doi.org/10.1016/j.shpsb.2003.10.002} {\bibfield
  {journal} {\bibinfo  {journal} {Studies in History and Philosophy of Science
  Part B: Studies in History and Philosophy of Modern Physics}\ }\textbf
  {\bibinfo {volume} {35}},\ \bibinfo {pages} {177} (\bibinfo {year} {2004})},\
  \Eprint {https://arxiv.org/abs/arXiv:quant-ph/0307139}
  {arXiv:quant-ph/0307139} \BibitemShut {NoStop}%
\bibitem [{\citenamefont {Abbott}\ \emph {et~al.}(2012)\citenamefont {Abbott},
  \citenamefont {Calude}, \citenamefont {Conder},\ and\ \citenamefont
  {Svozil}}]{2012-incomput-proofsCJ}%
  \BibitemOpen
  \bibfield  {author} {\bibinfo {author} {\bibfnamefont {A.~A.}\ \bibnamefont
  {Abbott}}, \bibinfo {author} {\bibfnamefont {C.~S.}\ \bibnamefont {Calude}},
  \bibinfo {author} {\bibfnamefont {J.}~\bibnamefont {Conder}},\ and\ \bibinfo
  {author} {\bibfnamefont {K.}~\bibnamefont {Svozil}},\ }\bibfield  {title}
  {\bibinfo {title} {Strong {K}ochen-{S}pecker theorem and incomputability of
  quantum randomness},\ }\href {https://doi.org/10.1103/PhysRevA.86.062109}
  {\bibfield  {journal} {\bibinfo  {journal} {Physical Review A}\ }\textbf
  {\bibinfo {volume} {86}},\ \bibinfo {pages} {062109} (\bibinfo {year}
  {2012})},\ \Eprint {https://arxiv.org/abs/arXiv:1207.2029} {arXiv:1207.2029}
  \BibitemShut {NoStop}%
\bibitem [{\citenamefont {Abbott}\ \emph
  {et~al.}(2014{\natexlab{a}})\citenamefont {Abbott}, \citenamefont {Calude},\
  and\ \citenamefont {Svozil}}]{PhysRevA.89.032109}%
  \BibitemOpen
  \bibfield  {author} {\bibinfo {author} {\bibfnamefont {A.~A.}\ \bibnamefont
  {Abbott}}, \bibinfo {author} {\bibfnamefont {C.~S.}\ \bibnamefont {Calude}},\
  and\ \bibinfo {author} {\bibfnamefont {K.}~\bibnamefont {Svozil}},\
  }\bibfield  {title} {\bibinfo {title} {Value-indefinite observables are
  almost everywhere},\ }\href {https://doi.org/10.1103/PhysRevA.89.032109}
  {\bibfield  {journal} {\bibinfo  {journal} {Physical Review A}\ }\textbf
  {\bibinfo {volume} {89}},\ \bibinfo {pages} {032109} (\bibinfo {year}
  {2014}{\natexlab{a}})},\ \Eprint {https://arxiv.org/abs/arXiv:1309.7188}
  {arXiv:1309.7188} \BibitemShut {NoStop}%
\bibitem [{\citenamefont {Jaynes}(2012)}]{jaynes}%
  \BibitemOpen
  \bibfield  {author} {\bibinfo {author} {\bibfnamefont {E.~T.}\ \bibnamefont
  {Jaynes}},\ }\href {https://doi.org/10.1017/CBO9780511790423} {\emph
  {\bibinfo {title} {Probability Theory: {T}he Logic Of Science}}}\ (\bibinfo
  {publisher} {Cambridge University Press},\ \bibinfo {address} {Cambridge},\
  \bibinfo {year} {2003,2012})\ \bibinfo {note} {edited by G. Larry
  Bretthorst}\BibitemShut {NoStop}%
\bibitem [{\citenamefont {Kleene}(1936)}]{Kleene1936}%
  \BibitemOpen
  \bibfield  {author} {\bibinfo {author} {\bibfnamefont {S.~C.}\ \bibnamefont
  {Kleene}},\ }\bibfield  {title} {\bibinfo {title} {General recursive
  functions of natural numbers},\ }\href {https://doi.org/10.1007/BF01565439}
  {\bibfield  {journal} {\bibinfo  {journal} {Mathematische Annalen}\ }\textbf
  {\bibinfo {volume} {112}},\ \bibinfo {pages} {727} (\bibinfo {year}
  {1936})}\BibitemShut {NoStop}%
\bibitem [{\citenamefont {Abbott}\ \emph
  {et~al.}(2014{\natexlab{b}})\citenamefont {Abbott}, \citenamefont {Calude},\
  and\ \citenamefont {Svozil}}]{Abbott:2010uq}%
  \BibitemOpen
  \bibfield  {author} {\bibinfo {author} {\bibfnamefont {A.~A.}\ \bibnamefont
  {Abbott}}, \bibinfo {author} {\bibfnamefont {C.~S.}\ \bibnamefont {Calude}},\
  and\ \bibinfo {author} {\bibfnamefont {K.}~\bibnamefont {Svozil}},\
  }\bibfield  {title} {\bibinfo {title} {A quantum random number generator
  certified by value indefiniteness},\ }\href
  {https://doi.org/10.1017/S0960129512000692} {\bibfield  {journal} {\bibinfo
  {journal} {Mathematical Structures in Computer Science}\ }\textbf {\bibinfo
  {volume} {24}},\ \bibinfo {pages} {e240303} (\bibinfo {year}
  {2014}{\natexlab{b}})},\ \Eprint {https://arxiv.org/abs/arXiv:1012.1960}
  {arXiv:1012.1960} \BibitemShut {NoStop}%
\bibitem [{\citenamefont {Svozil}(1999)}]{svozil-1999-haunted-qc}%
  \BibitemOpen
  \bibfield  {author} {\bibinfo {author} {\bibfnamefont {K.}~\bibnamefont
  {Svozil}},\ }\href {https://arxiv.org/abs/quant-ph/9907015} {\bibinfo {title}
  {``{H}aunted'' quantum contextuality}} (\bibinfo {year} {1999}),\ \Eprint
  {https://arxiv.org/abs/arXiv:quant-ph/9907015} {arXiv:quant-ph/9907015}
  \BibitemShut {NoStop}%
\bibitem [{\citenamefont {Svozil}(2009)}]{svozil:040102}%
  \BibitemOpen
  \bibfield  {author} {\bibinfo {author} {\bibfnamefont {K.}~\bibnamefont
  {Svozil}},\ }\bibfield  {title} {\bibinfo {title} {Proposed direct test of a
  certain type of noncontextuality in quantum mechanics},\ }\href
  {https://doi.org/10.1103/PhysRevA.80.040102} {\bibfield  {journal} {\bibinfo
  {journal} {Physical Review A}\ }\textbf {\bibinfo {volume} {80}},\ \bibinfo
  {eid} {040102} (\bibinfo {year} {2009})}\BibitemShut {NoStop}%
\bibitem [{\citenamefont {Griffiths}(2017)}]{Griffiths2017}%
  \BibitemOpen
  \bibfield  {author} {\bibinfo {author} {\bibfnamefont {R.~B.}\ \bibnamefont
  {Griffiths}},\ }\bibfield  {title} {\bibinfo {title} {What quantum
  measurements measure},\ }\bibfield  {journal} {\bibinfo  {journal} {Physical
  Review A}\ }\textbf {\bibinfo {volume} {96}},\ \href
  {https://doi.org/10.1103/physreva.96.032110} {10.1103/physreva.96.032110}
  (\bibinfo {year} {2017}),\ \Eprint {https://arxiv.org/abs/arXiv:1704.08725}
  {arXiv:1704.08725} \BibitemShut {NoStop}%
\bibitem [{\citenamefont {Griffiths}(2019)}]{Griffiths2019}%
  \BibitemOpen
  \bibfield  {author} {\bibinfo {author} {\bibfnamefont {R.~B.}\ \bibnamefont
  {Griffiths}},\ }\bibfield  {title} {\bibinfo {title} {Quantum measurements
  and contextuality},\ }\href {https://doi.org/10.1098/rsta.2019.0033}
  {\bibfield  {journal} {\bibinfo  {journal} {Philosophical Transactions of the
  Royal Society A: Mathematical, Physical and Engineering Sciences}\ }\textbf
  {\bibinfo {volume} {377}},\ \bibinfo {pages} {20190033} (\bibinfo {year}
  {2019})},\ \Eprint {https://arxiv.org/abs/arXiv:1902.05633}
  {arXiv:1902.05633} \BibitemShut {NoStop}%
\bibitem [{\citenamefont {Specker}(1990)}]{specker-ges}%
  \BibitemOpen
  \bibfield  {author} {\bibinfo {author} {\bibfnamefont {E.}~\bibnamefont
  {Specker}},\ }\href {https://doi.org/10.1007/978-3-0348-9259-9} {\emph
  {\bibinfo {title} {Selecta}}}\ (\bibinfo  {publisher} {Birkh{\"{a}}user
  Verlag},\ \bibinfo {address} {Basel},\ \bibinfo {year} {1990})\BibitemShut
  {NoStop}%
\bibitem [{\citenamefont {Stigler}(1980)}]{Stigler1980}%
  \BibitemOpen
  \bibfield  {author} {\bibinfo {author} {\bibfnamefont {S.~M.}\ \bibnamefont
  {Stigler}},\ }\bibfield  {title} {\bibinfo {title} {{S}tigler's law of
  eponymy},\ }\href {https://doi.org/10.1111/j.2164-0947.1980.tb02775.x}
  {\bibfield  {journal} {\bibinfo  {journal} {Transactions of the New York
  Academy of Sciences}\ }\textbf {\bibinfo {volume} {39}},\ \bibinfo {pages}
  {147} (\bibinfo {year} {1980})},\ \bibinfo {note} {in ``Science and social
  structure: a {F}estschrift for {R}obert {K}. {M}erton'', ed. by {T}homas {F}.
  {G}ieryn, reprinted in~\cite{Stigler-sott}}\BibitemShut {NoStop}%
\bibitem [{\citenamefont {Zierler}\ and\ \citenamefont
  {Schlessinger}(1975)}]{Zierler1975}%
  \BibitemOpen
  \bibfield  {author} {\bibinfo {author} {\bibfnamefont {N.}~\bibnamefont
  {Zierler}}\ and\ \bibinfo {author} {\bibfnamefont {M.}~\bibnamefont
  {Schlessinger}},\ }\bibfield  {title} {\bibinfo {title} {Boolean embeddings
  of orthomodular sets and quantum logic},\ }in\ \href
  {https://doi.org/10.1007/978-94-010-1795-4_14} {\emph {\bibinfo {booktitle}
  {The Logico-Algebraic Approach to Quantum Mechanics: Volume {I}: Historical
  Evolution}}},\ \bibinfo {editor} {edited by\ \bibinfo {editor} {\bibfnamefont
  {C.~A.}\ \bibnamefont {Hooker}}}\ (\bibinfo  {publisher} {Springer
  Netherlands},\ \bibinfo {address} {Dordrecht},\ \bibinfo {year} {1975})\ pp.\
  \bibinfo {pages} {247--262}\BibitemShut {NoStop}%
\bibitem [{\citenamefont {Stigler}(2002)}]{Stigler-sott}%
  \BibitemOpen
  \bibfield  {author} {\bibinfo {author} {\bibfnamefont {S.~M.}\ \bibnamefont
  {Stigler}},\ }\bibinfo {title} {Statistics on the table. {T}he history of
  statistical concepts and methods}\ (\bibinfo  {publisher} {Harvard University
  Press},\ \bibinfo {address} {Cambridge, MA, USA and London, England},\
  \bibinfo {year} {1999,2002})\ pp.\ \bibinfo {pages} {277--290}\BibitemShut
  {NoStop}%
\end{thebibliography}


%%%%%%%%%%%%%%%%%%%%%%%%%%%%%%%%%%%%%
%%%%%%%%%%%%%%%%%%%%%%%%%%%%%%%%%%%%%

\fi

\end{document}